\documentclass[10pt,journal]{IEEEtran}
\usepackage[dvipdfmx]{graphicx}
\usepackage{amsmath, amssymb, amsthm}
\usepackage{mathtools}
\usepackage{algorithm, algpseudocode}
\usepackage{xcolor}
\usepackage{float}
\usepackage{bbm}
\usepackage{cite}
\usepackage{hyperref}
\usepackage{enumerate}

\def\Bra#1{\left\langle#1\right|}
\def\Ket#1{\left|#1\right\rangle}
\newcommand{\bra}[1]{\langle{#1}|}
\newcommand{\ket}[1]{|{#1}\rangle}

\DeclareMathOperator{\rank}{rank}
\DeclareMathOperator{\tr}{Tr}
\DeclareMathOperator{\id}{id}
\DeclareMathOperator{\spn}{span}
\DeclareMathOperator{\supp}{supp}
\DeclareMathOperator*{\argmax}{argmax}

\theoremstyle{definition}
\newtheorem{theorem}{Theorem}
\newtheorem{lemma}[theorem]{Lemma}

\newtheorem{corollary}[theorem]{Corollary}
\newtheorem{proposition}[theorem]{Proposition}
\theoremstyle{definition}
\newtheorem{definition}[theorem]{Definition}
\newtheorem{example}{Example}
\newtheorem{implication}{Implication}
\theoremstyle{remark}
\newtheorem{conjecture}{Conjecture}
\newtheorem{remark}[conjecture]{Remark}

\begin{document}

\title{Quantum State Merging for Arbitrarily Small-Dimensional Systems}

\author{Hayata~Yamasaki and~Mio~Murao
  \thanks{This work was supported by Grant-in-Aid for JSPS Research Fellow and JSPS KAKENHI Grant Numbers 26330006, 15H01677, 16H01050, 17H01694, 18H04286, and 18J10192.}%
  \thanks{H.~Yamasaki is with the Department of Physics, Graduate School of Science, The University of Tokyo, Tokyo, Japan (email: \href{mailto:yamasaki@eve.phys.s.u-tokyo.ac.jp}{yamasaki@eve.phys.s.u-tokyo.ac.jp}).}%
  \thanks{M.~Murao is with the Department of Physics, Graduate School of Science,
    The University of Tokyo, Tokyo, Japan (email: \href{mailto:murao@phys.s.u-tokyo.ac.jp}{murao@phys.s.u-tokyo.ac.jp}).}%
}%

\maketitle

\date{\today}

\begin{abstract}
  Recent advances in quantum technology facilitate the realization of information processing using quantum computers at least on the small and intermediate scales of up to several dozens of qubits. We investigate entanglement cost required for one-shot quantum state merging, aiming at quantum state transformation on these scales. In contrast to existing coding algorithms achieving nearly optimal approximate quantum state merging on a large scale, we construct algorithms for exact quantum state merging so that the algorithms are applicable to any given state of an arbitrarily small-dimensional system. In the algorithms, entanglement cost can be reduced depending on a structure of the given state derived from the Koashi-Imoto decomposition. We also provide improved converse bounds for exact quantum state merging achievable for qubits but not necessarily achievable in general. As for approximate quantum state merging, we obtain algorithms and improved converse bounds by applying smoothing to those for exact state merging. Our results are applicable to distributed quantum information processing and multipartite entanglement transformation on small and intermediate scales.
\end{abstract}


\section{\label{sec:intro}Introduction}

\IEEEPARstart{T}{he era} of small- and intermediate-scale quantum computers of up to several dozens of qubits is approaching due to advances in quantum technology.
There exists, however, technical difficulty in increasing the number of low-noise qubits built in one quantum device~\cite{P4}.
For further scaling up, distributed quantum information processing using multiple quantum devices connected by a network for quantum communication is considered to be promising~\cite{V3,C9}.
Aimed at efficient quantum information processing,
coding algorithms for quantum communication tasks in such a distributed setting should be designed to be suitable for transferring quantum states on these small and intermediate scales.

Quantum state merging~\cite{H3,H4} is a task playing crucial roles in distributed quantum information processing~\cite{W8,W9,W10} and multipartite entanglement transformations~\cite{A3,D8,Y8,D9,S4}.
Originally, state merging, or state redistribution~\cite{D2,D3} as a generalized task including state merging,
was introduced in the context of quantum Shannon theory, and it has applied to the analyses of various tasks in quantum Shannon theory such as derivation of a capacity of noisy quantum channels~\cite{D4,A7,H5,A2,H11,A8,P3,W5}.
In the task of state merging formulated in the original paper~\cite{H3} using the framework of local operations and classical communication (LOCC),
two spatially separated parties $A$ and $B$ initially share an entangled resource state and are given $n$ mixed states whose purification with reference $R$ is represented as ${\left(\Ket{\psi}^{RAB}\right)}^{\otimes n}$, where $A$ and $B$ knows classical description of $\Ket{\psi}^{RAB}$.
The goal of the task is to asymptotically transfer $A$'s part of $\Ket{\psi}^{RAB}$ to $B$ and obtain $\Ket{\psi}^{RB'B}$,
keeping coherence between $B$ and $R$, by LOCC assisted by shared entanglement within an error in fidelity approaching to zero as $n \rightarrow \infty$.
State merging can also be regarded as an analogue of source coding with decoder's side information in classical information theory established by Slepian and Wolf~\cite{S7}, which aims at compressing $A$'s classical message exploiting $B$'s side information on the message.

\begin{figure}[t]
  \centering
  \includegraphics[width=3.5in]{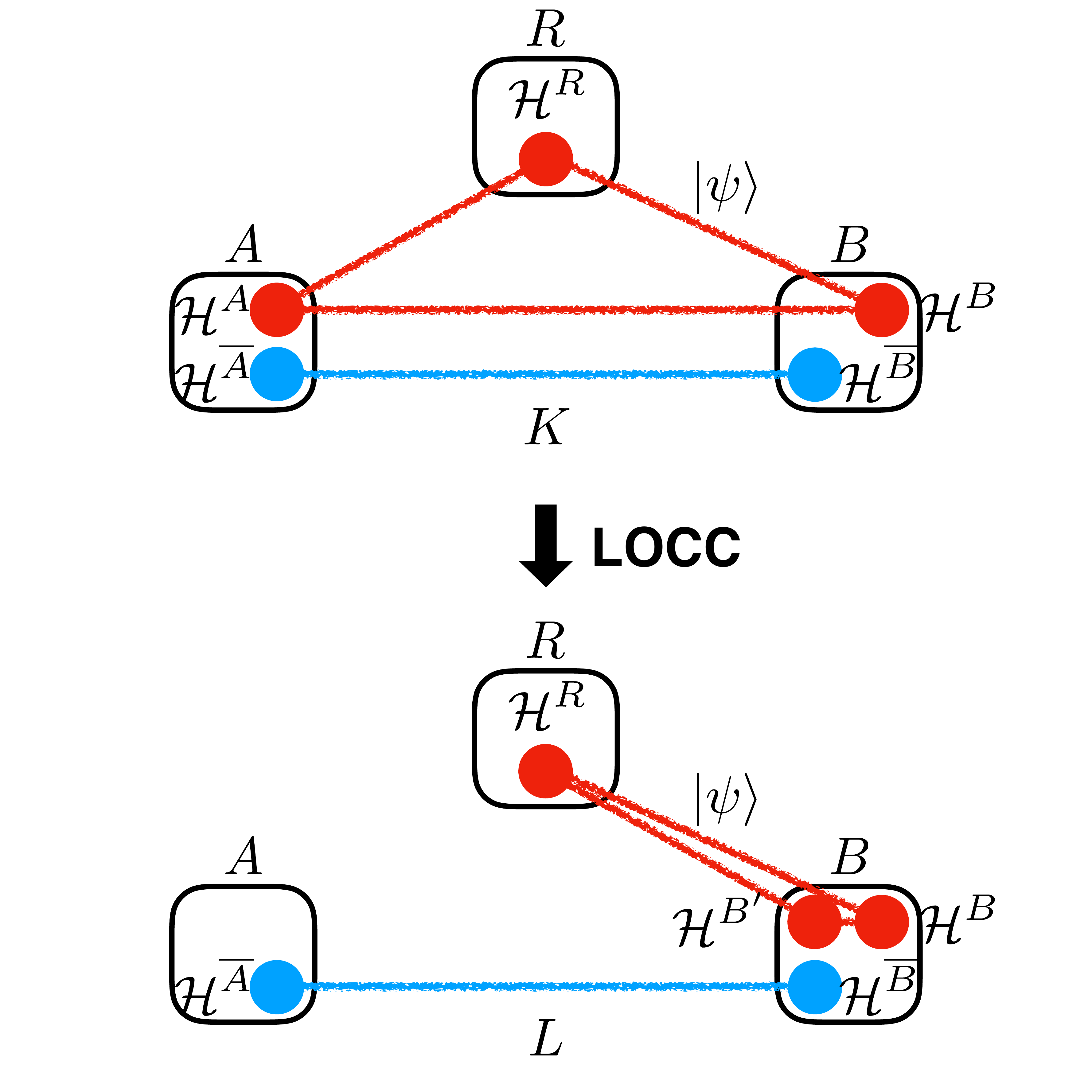}
  \caption{\label{fig:merge} Exact state merging of a given state $\Ket{\psi}^{RAB}$ denoted by the red circles. Parties $A$ and $B$ perform LOCC assisted by a maximally entanglement resource state $\Ket{\Phi_K^+}^{\overline{AB}}$ with the Schmidt rank $K$ denoted by the top blue circles to transfer $A$'s part of $\Ket{\psi}^{RAB}$ to $B$ and obtain $\Ket{\psi}^{RB'B}$ while $\Ket{\Phi_L^+}^{\overline{AB}}$ with the Schmidt rank $L$ denoted by the bottom blue circles is also obtained.}
\end{figure}

It is an essential feature of quantum state merging that the parties may exploit classical description of $\Ket{\psi}^{RAB}$ for reducing the amount of entanglement consumption, or \textit{entanglement cost}, required for an algorithm achieving state merging of $\Ket{\psi}^{RAB}$.
Without classical description of $\Ket{\psi}^{RAB}$, there exists a trivial algorithm achieving state merging by quantum teleportation~\cite{B5} to transfer $A$'s part of $\Ket{\psi}^{RAB}$ from $A$ to $B$.
In contrast, entanglement cost in state merging can be reduced compared to quantum teleportation and even be negative when the algorithm provides a net gain of shared entanglement.

While this type of asymptotic scenarios are well-established in quantum Shannon theory, there have also been studied zero-error scenarios~\cite{G5}, which are originally established in a classical setting by Shannon~\cite{S6} and first introduced into a quantum setting in Ref.~\cite{M2}.
In the zero-error scenarios of classical source coding with decoder's side information, optimal zero-error code design is proven to be $NP$-hard~\cite{K7}.
However, in classical coding theory, explicit construction of zero-error coding algorithms such as Shannon coding~\cite{S9} and Huffman coding~\cite{H9}, if not necessarily optimal, establishes a foundation of theoretical analyses as well as practical applications.
In this direction, explicit zero-error coding algorithms for classical source coding with decoder's side information are shown in Refs.~\cite{K7,W7,J2,Y11,Z2,M3,M4}.

Aside from this regime where infinitely many copies of $\Ket{\psi}^{RAB}$ are given, another regime is the one-shot regime where only a single copy is given.
The scenarios in the one-shot regime can also be classified into two scenarios: one is an exact scenario with zero error, and the other is an approximate scenario in which a nonzero error is tolerated for reducing entanglement cost.
Analysis in the one-shot regime clarifies the structure of algorithms achieving the task at a single-copy level and is more relevant to practical situations such as distributed quantum information processing.

However, the existing algorithms for one-shot quantum state merging or redistribution~\cite{B9,Y9,B12,D7,D6,H10,B10,D5,M,N3,A4,A5} achieve near optimality only on a large scale relevant to \textit{one-shot quantum information theory} where the smooth conditional min- and max-entropies~\cite{R2,T5} are used to evaluate entanglement cost.
These algorithms also need a nonzero approximation error in fidelity, since the vital techniques for these algorithms, namely, one-shot decoupling~\cite{D6} and the convex-split lemma~\cite{A4}, require errors.
As higher fidelity is pursued in state merging of a fixed single copy of $\Ket{\psi}^{RAB}$, entanglement cost required for the algorithms diverges to infinity. Hence, there always exists a region of error close to zero where the algorithms do not contribute to reducing the entanglement cost.
Moreover,
in cases where $A$'s system size for $\Ket{\psi}^{RAB}$ is as small as up to a few dozens of qubits,
the algorithms require more entanglement cost than quantum teleportation, even if the error tolerance is reasonably large (see Remark~\ref{remark:usefulness} in Sec.~\ref{sec:merge} for more discussion).
In this sense, strategies in state merging to exploit the classical description of $\Ket{\psi}^{RAB}$ for reducing entanglement cost have \textit{not} yet been established for arbitrarily small-dimensional systems or arbitrarily high fidelity.

In this paper, we explicitly construct algorithms for one-shot state merging
which have the following features:
\begin{enumerate}
  \item Applicable to any state including small- and intermediate-scale states;
  \item Fulfilling arbitrarily high fidelity requirement including zero error;
  \item Retaining the essential feature of state merging, that is, exploiting classical description of $\Ket{\psi}^{RAB}$ for reducing entanglement cost.
\end{enumerate}%
The tasks of one-shot state merging investigated in this paper are achieved exactly, that is, without approximation, which we call \textit{exact state merging} (Fig.~\ref{fig:merge}).
Entanglement cost of our algorithms for exact state merging is not larger than, and can be strictly smaller than,
the optimal entanglement cost in its inverse task, exact state splitting, depending on a decomposition of $\Ket{\psi}^{RAB}$ referred to in Ref.~\cite{W4} as the Koashi-Imoto decomposition~\cite{W4,K3,H6,K5}.
We show multiple examples of states including those relevant to distributed quantum information processing where our algorithms for exact state merging can reduce entanglement cost since these states have \textit{nontrivial} Koashi-Imoto decomposition.
In the same way as the asymptotic scenarios, the entanglement cost of our algorithm can even be negative.
In addition to providing achievability bounds, we improve the existing converse bound~\cite{B9} of entanglement cost in exact state merging given in terms of the conditional max-entropy and show that our converse bound is achievable when the state to be merged is represented by qubits.
By means of \textit{smoothing}~\cite{R2,T5},
we also extend these results to \textit{approximate state merging}, where arbitrarily small approximation error in fidelity is allowed so that the entanglement cost can further be reduced compared to exact state merging, and our converse bound of entanglement cost in approximate state merging improves the existing converse bound~\cite{B10}.

This paper is organized as follows.
In Sec.~\ref{sec:preliminaries}, we introduce definitions of exact state merging and provide a summary of the Koashi-Imoto decomposition.
In Sec.~\ref{sec:result}, we present our main results: Theorems~\ref{thm:merge} and~\ref{thm:merge_without_catalyst} for achievability of exact state merging and Theorem~\ref{thm:new} for converse.
Extension of these results to approximate state merging is presented in Appendix~\ref{sec:approximate}.
Implications are discussed in Sec.~\ref{sec:examples}.
Our conclusion is given in Sec.~\ref{sec:conclusion}.
Exact state splitting is also analyzed in Appendix~\ref{sec:split}, where Theorem~\ref{thm:split} yields the optimal entanglement cost in exact state splitting.

\section{\label{sec:preliminaries}Preliminaries}
In this section, after presenting our notations in Sec.~\ref{sec:notations}, we define exact state merging in Sec.~\ref{sec:def_merge}.
Then, we introduce the Koashi-Imoto decomposition in Sec.~\ref{sec:koashi_imoto_decomposition}.

\subsection{\label{sec:notations}Notations}
We represent a system indexed by $X$ as a Hilbert space denoted by $\mathcal{H}^X$.  The set of density operators on $\mathcal{H}^X$ is denoted by $\mathcal{D}\left(\mathcal{H}^X\right)$. The set of bounded operators on $\mathcal{H}^X$ is denoted by $\mathcal{B}\left(\mathcal{H}^X\right)$.   Superscripts of an operator or a vector represent the indices of the corresponding Hilbert spaces, \textit{e.g.}, $\psi^{RA} \in \mathcal{D} (\mathcal{H}^R\otimes\mathcal{H}^{A})$ for a mixed state and $\Ket{\psi}^{R A B} \in\mathcal{H}^R\otimes\mathcal{H}^A\otimes\mathcal{H}^{B}$ for a pure state.
We may write an operator representing a pure state as $\psi^{RAB}\coloneqq\Ket{\psi}\Bra{\psi}^{R A B}$. A reduced state may be represented by superscripts if obvious, such as $\psi^{RA}\coloneqq\tr_{B}\psi^{RAB}$.
The identity operator and the identity map on $\mathcal{H}^X$ are denoted by $\mathbbm{1}^X$ and $\id^X$, respectively.
In particular, to explicitly show the dimension of an identity operator, we may use subscripts, \textit{e.g.} the identity operator on $\mathcal{H}^R$ of dimension $D$ may be denoted by $\mathbbm{1}_D^R$.

\subsection{\label{sec:def_merge}Definition of exact state merging}
Exact state merging involves three parties $A$, $B$, and $R$,
where $R$ is a reference to consider purification.
Let $A$ have $\mathcal{H}^A$ and $\mathcal{H}^{\overline{A}}$, $B$ have $\mathcal{H}^B$, $\mathcal{H}^{B'}$, and $\mathcal{H}^{\overline{B}}$, and $R$ have $\mathcal{H}^R$,
where $\dim\mathcal{H}^{A}=\dim\mathcal{H}^{B'}$.
We assume that the parties $A$ and $B$ can freely perform LOCC assisted by a maximally entangled resource state on $\mathcal{H}^{\overline{A}}\otimes\mathcal{H}^{\overline{B}}$ initially shared between $A$ and $B$.
Regarding a formal definition of LOCC, refer to Ref.~\cite{C7}.
Note that $A$ and $B$ cannot perform any operation on $\mathcal{H}^R$.

We define the task of exact state merging as illustrated in Fig.~\ref{fig:merge}.
Initially, $A$ and $B$ are given a possibly mixed state of $\mathcal{H}^A\otimes\mathcal{H}^B$ whose purification is represented by $\Ket{\psi}^{RAB}$, where $A$ and $B$ knows classical description of $\Ket{\psi}^{RAB}$.
Exact state merging of $\Ket{\psi}^{RAB}$ is a task for $A$ and $B$ to exactly transfer $A$'s part of $\Ket{\psi}^{RAB}$ from $A$ to $B$ and obtain $\Ket{\psi}^{RB'B}$.
The given state $\Ket{\psi}^{RAB}$ may have entanglement between $A$ and $B$,
and hence, $A$ and $B$ may also be able to distill this entanglement.
Let $K$ denote the Schmidt rank of an initial resource state
\[
  \Ket{\Phi^+_K}^{\overline{A}\overline{B}}\coloneqq\frac{1}{\sqrt{K}}\sum_{l=0}^{K-1}\Ket{l}^{\overline{A}}\otimes\Ket{l}^{\overline{B}}
\]
shared between $A$ and $B$ before performing exact state merging.
After completing exact state merging, $A$ and $B$ may obtain a final resource state
\[
  \Ket{\Phi^+_L}^{\overline{A}\overline{B}}\coloneqq\frac{1}{\sqrt{L}}\sum_{l=0}^{L-1}\Ket{l}^{\overline{A}}\otimes\Ket{l}^{\overline{B}}
\]
with the Schmidt rank $L$
to be used in the future.
If $\log_2 K - \log_2 L\geqq 0$,
$\log_2 K - \log_2 L$ is regarded as the amount of net entanglement consumption in exact state merging,
and otherwise $\log_2 L - \log_2 K$ is regarded as the amount of net entanglement gain.
In cases where $\log_2 K > 0$ and $\log_2 L > 0$,
a part of entanglement in the initial resource state is interpreted to be used catalytically,
where an initial resource state with larger $\log_2 K$ may be used to decrease $\log_2 K - \log_2 L$.
We call this setting the \textit{catalytic} setting.
On the other hand, simply minimizing the amount of entanglement of the initial resource state may also be useful especially in the one-shot regime.
Thus, we also consider another setting of fixing $\log_2 L = 0$ as a variant of exact state merging, where the catalytic use of shared entanglement is forbidden.
We call such a task \textit{non-catalytic} exact state merging.

\begin{definition}
\label{def:merging}
    \textit{Exact state merging.}
    Exact state merging of a purified given state $\Ket{\psi}^{RAB}$ is a task for parties $A$ and $B$ to achieve a transformation
    \[
        \begin{split}
            \id^R\otimes\mathcal{M}\left({\psi}^{RAB}\otimes{\Phi^+_K}^{\overline{A}\overline{B}}\right)
            ={\psi}^{RB'B}\otimes{\Phi^+_L}^{\overline{A}\overline{B}}
        \end{split}
    \]
    by an LOCC map $\mathcal{M}:\mathcal{B}\left(\mathcal{H}^A\otimes\mathcal{H}^B\otimes\mathcal{H}^{\overline{A}}\otimes\mathcal{H}^{\overline{B}}\right)\to\mathcal{B}\left(\mathcal{H}^{B'}\otimes\mathcal{H}^B\otimes\mathcal{H}^{\overline{A}}\otimes\mathcal{H}^{\overline{B}}\right)$, which can be constructed depending on the classical description of $\Ket{\psi}^{RAB}$.
    The definition of \textit{non-catalytic} exact state merging is also obtained by setting $\log_2 L=0$ in the above definition.
\end{definition}

Entanglement cost of an algorithm for exact state merging in the catalytic setting is defined as $\log_2 K-\log_2 L$, and that for non-catalytic exact state merging is defined as $\log_2 K$.
The minimal entanglement cost among all the algorithms for exact state merging of $\Ket{\psi}^{RAB}$ may be simply referred to as entanglement cost in exact state merging of $\Ket{\psi}^{RAB}$.
If $\log_2 K\geqq\log_2 \dim\mathcal{H}^A$, there exists a trivial algorithm for exact state merging by quantum teleportation to transfer $\psi^{A}$ from $A$ to $B$.
Our results given in Sec.~\ref{sec:result} provide algorithms at less entanglement cost using the classical description of $\Ket{\psi}^{RAB}$.

There exist following tasks achievable at the same entanglement cost using the same algorithm as those in exact state merging of a given state $\Ket{\psi}^{RAB}$, as shown in Appendix~\ref{sec:equivalence}.
  Consider the Schmidt decomposition of $\Ket{\psi}^{RAB}$ with respect to bipartition between $\mathcal{H}^R$ and $\mathcal{H}^{A}\otimes\mathcal{H}^{B}$
  \begin{equation}
    \label{eq:schmidt}
    \Ket{\psi}^{RAB}=\sum_{l=0}^{D-1} \sqrt{\lambda_l}\Ket{l}^R\otimes\Ket{\psi_l}^{AB},
  \end{equation}
  where $D$ is the Schmidt rank, and $\lambda_l>0$ for each $l\in\{0,\ldots,D-1\}$.
  Then, entanglement cost in exact state merging of $\Ket{\psi}^{RAB}$ equals to that of a maximally entangled state $\Ket{\Phi_D^+\left(\psi\right)}^{RAB}$ with Schmidt rank $D$ corresponding to $\Ket{\psi}^{RAB}$
  \begin{equation}
    \label{eq:max}
    \Ket{\Phi_D^+\left(\psi\right)}^{RAB}\coloneqq\sum_{l=0}^{D-1} \frac{1}{\sqrt{D}}\Ket{l}^R\otimes\Ket{\psi_l}^{AB},
  \end{equation}
  where the Schmidt basis on the right-hand side is the same as that in Eq.~\eqref{eq:schmidt}.
  This equivalence is used for simplifying our analysis in Sec.~\ref{sec:converse}.

  This equivalence also implies that entanglement cost in exact state merging of $\Ket{\psi}^{RAB}$ is the same as that required for merging arbitrary bipartite states shared between $A$ and $B$ on a subspace of $\mathcal{H}^A\otimes\mathcal{H}^B$ spanned by the Schmidt-basis states ${\left\{\Ket{\psi_l}^{AB}\right\}}_l$ corresponding to nonzero Schmidt coefficients in Eq.~\eqref{eq:schmidt}.
  The equivalence between considering the maximally entangled state with $R$ in Eq.~\eqref{eq:max} and considering arbitrary bipartite states on the corresponding subspace is also known as the relative state method~\cite{P3}.
  Note that in general, entanglement cost in exact state merging of $\Ket{\psi}^{RAB}$ is different from that required for merging arbitrary bipartite states given from an ensemble ${\left\{p\left(l\right),\Ket{\psi_l}^{AB}\right\}}_l$ for a probability distribution $p\left(l\right)$, since coherence of arbitrary superposition of ${\left\{\Ket{\psi_l}^{AB}\right\}}_l$ has to be kept in state merging of $\Ket{\psi}^{RAB}$.

\subsection{\label{sec:koashi_imoto_decomposition}Koashi-Imoto decomposition}

We summarize the \textit{Koashi-Imoto decomposition}~\cite{K3,H6,K5,W4}.
The Koashi-Imoto decomposition is first introduced in Ref.~\cite{K3} to characterize a completely positive and trace-preserving (CPTP) map leaving any state in a given set invariant.
Reference~\cite{W4} extends the Koashi-Imoto decomposition to that defined for a given tripartite pure state $\Ket{\psi}^{RAB}$.
The Koashi-Imoto decomposition of $\Ket{\psi}^{RAB}$ is obtained using a set of $A$'s states which can be \textit{steered} through the bipartite reduce state $\psi^{RA}$, that is, the set of states of $\mathcal{H}^A$ which can be prepared by performing a measurement of $\psi^{RA}$ on $\mathcal{H}^R$ and post-selecting an outcome.
Using an arbitrary positive semidefinite operator $\Lambda^R$,
this set of states is denoted by
\begin{equation}
  \label{eq:psi_lambda}
  \begin{split}
    S_\psi^{A|R}&\coloneqq{\left\{\psi^A\left(\Lambda^R\right):\Lambda^R\geqq 0\right\}},\\
    \psi^A\left(\Lambda^R\right)&\coloneqq\frac{\tr_R \left[\left(\Lambda^R\otimes\mathbbm{1}^A\right)\psi^{RA}\right]}{\tr \left[\left(\Lambda^R\otimes\mathbbm{1}^A\right)\psi^{RA}\right]},
  \end{split}
\end{equation}
where the post-selected outcome of a measurement of $\psi^{RA}$ on $\mathcal{H}^R$ corresponds to $\Lambda^R$.
Using this notation, the Koashi-Imoto decomposition of a tripartite pure state is shown as follows.
Note that due to the second condition in the following lemma, the Koashi-Imoto decomposition is \textit{uniquely} determined, corresponding to the decomposition said to be maximal in Ref.~\cite{K3}.

\begin{lemma}
\label{lem:koashi_imoto_decomposition_tripartite}
  (Lemma~11 in Ref.~\cite{W4})
  \textit{Koashi-Imoto decomposition of a tripartite pure state.}
  Given any tripartite pure state $\Ket{\psi}^{RAB}$,
  there exists a unique decomposition of $\mathcal{H}^A$ in the form of
  \begin{equation}
    \label{eq:notation_space_a}
    \mathcal{H}^A=\bigoplus_{j=0}^{J-1}\mathcal{H}^{a_j^L}\otimes\mathcal{H}^{a_j^R}
  \end{equation}
  satisfying the following two conditions:
  \begin{enumerate}
    \item The support $\supp\left(\psi^B\right)$ of $\psi^B$ is decomposed into
      \begin{equation}
        \label{eq:notation_space_b}
        \supp\left(\psi^B\right)=\bigoplus_{j=0}^{J-1}\mathcal{H}^{b_j^L}\otimes\mathcal{H}^{b_j^R},
      \end{equation}
      and $\Ket{\psi}^{RAB}$ is decomposed into
      \begin{equation}
        \label{eq:notation_state}
        \Ket{\psi}^{RAB}=\bigoplus_{j=0}^{J-1}\sqrt{p\left(j\right)}\Ket{\omega_j}^{a_j^L b_j^L}\otimes\Ket{\phi_j}^{R a_j^R b_j^R},
      \end{equation}
      where $p\left(j\right)$ is a probability distribution;
    \item For any CPTP map $\mathcal{T}:\mathcal{B}\left(\mathcal{H}^A\right)\to\mathcal{B}\left(\mathcal{H}^A\right)$,
      if $\mathcal{T}$ leaves $\psi^A\left(\Lambda^R\right)\in S_\psi^{A|R}$ defined as Eq.~\eqref{eq:psi_lambda} invariant for any $\Lambda^R\geqq 0$, that is, $\mathcal{T}\left(\psi^A\left(\Lambda^R\right)\right)=\psi^A\left(\Lambda^R\right)$,
      then any isometry $U_\mathcal{T}$ from $\mathcal{H}^A$ to $\mathcal{H}^A\otimes\mathcal{H}^{E}$ for $\mathcal{T}$'s Stinespring dilation $\mathcal{T}(\rho)=\tr_{E}U_\mathcal{T}\rho U_\mathcal{T}^\dag$ is decomposed into $U_\mathcal{T}=\bigoplus_{j=0}^{J-1} U_j^{a_j^L}\otimes\mathbbm{1}^{a_j^R}$,
      where for each $j\in\{0,\ldots,J-1\}$, $U_j^{a_j^L}$ is an isometry from $\mathcal{H}^{a_j^L}$ to $\mathcal{H}^{a_j^L}\otimes\mathcal{H}^{E}$ satisfying $\tr_{E}U_\mathcal{T} \omega_j^{a_j^L} U_\mathcal{T}^\dag = \omega_j^{a_j^L}$.
  \end{enumerate}
\end{lemma}

To obtain the Koashi-Imoto decomposition of a given pure state $\Ket{\psi}^{RAB}$, we can use an algorithm shown in Ref.~\cite{K3}, as demonstrated in terms of our notations in Appendix~\ref{sec:koashi_imoto_decomposition_algorithm}.

\section{\label{sec:result}Main results}
In this section, we first provide an algorithm achieving exact state merging and evaluate the achievability bound of entanglement cost in Sec.~\ref{sec:merge}. Then, we also analyze the converse bound of entanglement cost in exact state merging in Sec.~\ref{sec:converse}.
Extension of these results on exact state merging to approximate state merging is presented in Appendix~\ref{sec:approximate}.

\subsection{\label{sec:merge}Achievability bound for exact state merging applicable to arbitrarily small-dimensional systems}
We provide algorithms for exact state merging applicable to any state of an arbitrarily small-dimensional system,
using the Koashi-Imoto decomposition introduced in Sec.~\ref{sec:koashi_imoto_decomposition}.
Given any pure state $\Ket{\psi}^{RAB}$,
Lemma~\ref{lem:koashi_imoto_decomposition_tripartite} yields the unique decomposition of $\mathcal{H}^A$ and $\supp\left(\psi^B\right)$ shown in Eqs.~\eqref{eq:notation_space_a} and~\eqref{eq:notation_space_b}, respectively, and the unique decomposition of $\Ket{\psi}^{RAB}$ shown in Eq.~\eqref{eq:notation_state}.
Also, for each $j\in\{0,\ldots,J-1\}$,
we write the reduced state of $\Ket{\omega_j}^{a_j^L b_j^L}$ on $\mathcal{H}^{a_j^L}$ as
\begin{equation}
  \label{eq:notation_lambda}
  \begin{split}
    \omega_j^{a_j^L}\coloneqq\tr_{b_j^L}\Ket{\omega_j}\Bra{\omega_j}^{a_j^L b_j^L}
    =\sum_l\lambda_l^{a_j^L}\Ket{l}\Bra{l},
  \end{split}
\end{equation}
where $l\in\left\{0,\ldots,\rank\omega_j^{a_j^L}-1\right\}$, the right-hand side represents the spectral decomposition,
and we let $\lambda^{a_j^L}_0$ denote the largest eigenvalue of $\omega_j^{a_j^L}$.
Using the Koashi-Imoto decomposition,
we provide an algorithm for exact state merging, which yields the following theorem.

\begin{theorem}
\label{thm:merge}
    \textit{An achievability bound of entanglement cost in exact state merging applicable to arbitrarily small-dimensional systems.}
    Given any pure state $\Ket{\psi}^{RAB}$ and
    any $\delta > 0$,
    there exists an algorithm for exact state merging of $\Ket{\psi}^{RAB}$ achieving
    \begin{equation}
        \label{eq:merge_cost}
        \log_2 K-\log_2 L \leqq \max_{j}\left\{\log_2\left(\lambda_0^{a_j^L}\dim\mathcal{H}^{a_j^R}\right)\right\} + \delta,
    \end{equation}
    where the notations are the same as those in Eqs.~\eqref{eq:notation_space_a},~\eqref{eq:notation_space_b},~\eqref{eq:notation_state}, and~\eqref{eq:notation_lambda}.
\end{theorem}

As for non-catalytic exact state merging, the entanglement cost $\log_2 K$ of the initial resource state can be reduced compared to $\log_2 K$ required for the algorithm in the catalytic setting in Theorem~\ref{thm:merge}.
Note that, however, $\log_2 K$ for non-catalytic exact state merging may be more than the net entanglement cost $\log_2 K-\log_2 L$ required for the algorithm in the catalytic setting in Theorem~\ref{thm:merge}.

\begin{theorem}
\label{thm:merge_without_catalyst}
    \textit{An achievability bound of entanglement cost in non-catalytic exact state merging applicable to arbitrarily small-dimensional systems.}
    Given any pure state $\Ket{\psi}^{RAB}$,
    there exists an algorithm for non-catalytic exact state merging of $\Ket{\psi}^{RAB}$ achieving
    \begin{equation}
        \label{eq:merge_without_catalyst_cost}
        \log_2 K \leqq \max_{j}\left\{\log_2\left\lceil\lambda_0^{a_j^L}\dim\mathcal{H}^{a_j^R}\right\rceil\right\},
    \end{equation}
    where $\lceil{}\cdots{}\rceil$ is the ceiling function,
    and the other notations are the same as those in Theorem~\ref{thm:merge}.
\end{theorem}

\begin{IEEEproof}[Proof of Theorem~\ref{thm:merge}]
    We construct an algorithm for exact state merging of $\Ket{\psi}^{RAB}$ achieving Inequality~\eqref{eq:merge_cost}.
    We define
    \begin{align*}
      j_0&\coloneqq\argmax_{j}\left\{\log_2\left(\lambda_0^{a_j^L}\dim\mathcal{H}^{a_j^R}\right)\right\},\\
      D^{a_j^R}&\coloneqq\dim\mathcal{H}^{a_j^R}\quad
        \text{for each $j\in\left\{0,\ldots,J-1\right\}$.}
    \end{align*}
    We may omit identity operators, such as $\mathbbm{1}^R$, in the following for brevity.

    Our algorithm uses the following tensor-product form of the Koashi-Imoto decomposition of $\Ket{\psi}^{RAB}$, which is equivalent to that shown in Lemma~\ref{lem:koashi_imoto_decomposition_tripartite} as well as Eqs.~\eqref{eq:notation_space_a},~\eqref{eq:notation_space_b}, and~\eqref{eq:notation_state}.
    Given the Koashi-Imoto decomposition of $\Ket{\psi}^{RAB}$ in the form of Eq.~\eqref{eq:notation_state}, introducing auxiliary systems $\mathcal{H}^{a_0}$ and $\mathcal{H}^{b_0}$, we can also write this decomposition as
    \begin{equation}
      \label{eq:ki_tripartite_isometry}
      \begin{split}
        &\left(\mathbbm{1}^R\otimes U^A \otimes U^B\right) \Ket{\psi}^{RAB}\\
        &=\sum_{j=0}^{J-1}\sqrt{p\left(j\right)}\Ket{j}^{a_0}\otimes\Ket{j}^{b_0}\otimes\Ket{\omega_j}^{a^L b^L}\otimes\Ket{\phi_j}^{R a^R b^R},
      \end{split}
    \end{equation}
    where $\mathcal{H}^{a_0}$, $\mathcal{H}^{b_0}$, $\mathcal{H}^{a^L}$, $\mathcal{H}^{b^L}$, $\mathcal{H}^{a^R}$, and $\mathcal{H}^{b^R}$ satisfy
    \begin{align*}
      \dim\mathcal{H}^{a_0}&=J,\\
      \dim\mathcal{H}^{b_0}&=J,\\
      \dim\mathcal{H}^{a^L}&=\max_j\left\{\dim\mathcal{H}^{a_j^L}\right\},\\
      \dim\mathcal{H}^{b^L}&=\max_j\left\{\dim\mathcal{H}^{b_j^L}\right\},\\
      \dim\mathcal{H}^{a^R}&=\max_j\left\{\dim\mathcal{H}^{a_j^R}\right\},\\
      \dim\mathcal{H}^{b^R}&=\max_j\left\{\dim\mathcal{H}^{b_j^R}\right\},
    \end{align*}
    $U^A$ is an isometry from $\mathcal{H}^A$ to $\mathcal{H}^{a_0}\otimes\mathcal{H}^{a^L}\otimes\mathcal{H}^{a^R}$,
    $U^B$ is an isometry from $\mathcal{H}^B$ to $\mathcal{H}^{b_0}\otimes\mathcal{H}^{b^L}\otimes\mathcal{H}^{b^R}$,
    and ${\{\Ket{j}^{a_0}:j=0,\ldots,J-1\}}$ and ${\{\Ket{j}^{b_0}:j=0,\ldots,J-1\}}$ are the computational basis of $\mathcal{H}^{a_0}$ and $\mathcal{H}^{b_0}$, respectively.
    In the same way as stressed in Ref.~\cite{K3},
    information on $\psi^A$ is encoded in three parts of the Koashi-Imoto decomposition in Eq.~\eqref{eq:ki_tripartite_isometry}, namely, $\mathcal{H}^{a_0}$, $\mathcal{H}^{a^R}$, and $\mathcal{H}^{a^L}$, which can be regarded as the classical part, the nonclassical (quantum) part, and the redundant part, respectively.
    In the rest of the proof, we first present the following three subprocesses:
    \begin{enumerate}
      \item Entanglement distillation from the \textit{redundant} part;
      \item Quantum teleportation to transfer the \textit{quantum} part;
      \item Coherently merging the \textit{classical} part by a measurement.
    \end{enumerate}
    Then, we show a procedure for combining these three subprocesses, using controlled measurements and controlled isometries, which are controlled by states of $\mathcal{H}^{a_0}$ and $\mathcal{H}^{b_0}$.

    \textit{Subprocess~1: Entanglement distillation from the redundant part.}
    Due to the continuity of $\log_2$, there exists a rational number $\tilde{\lambda}_0^{a_{j_0}^L}\in\mathbb{Q}$, where $\mathbb{Q}$ denotes the set of rational numbers, such that
    \begin{align*}
      &\log_2\left(\lambda_{0}^{a_{j_0}^L}D^{a^R_{j_0}}\right)\\
      &\leqq\log_2\left(\tilde{\lambda}_{0}^{a_{j_0}^L}D^{a^R_{j_0}}\right)\\
      &\leqq\log_2\left(\lambda_{0}^{a_{j_0}^L}D^{a^R_{j_0}}\right)+\delta.
    \end{align*}
    Thus, for any $j\in\left\{0,\ldots,J-1\right\}$, it holds that
    \begin{align*}
        \lambda_{0}^{a_{j}^L}D^{a^R_j}
        \leqq\lambda_{0}^{a_{j_0}^L}D^{a^R_{j_0}}
        \leqq\tilde{\lambda}_{0}^{a_{j_0}^L}D^{a^R_{j_0}}.
    \end{align*}
    Hence, we have
    \[
        \lambda_0^{a_{j}^L}\leqq\frac{D^{a^R_{j_0}}}{D^{a^R_j}}{\tilde{\lambda}_0^{a_{j_0}^L}},
    \]
    and since $\tilde{\lambda}_0^{a_{j_0}^L}\in\mathbb{Q}$, there exist integers $K_j$ and $L_j$ such that the right-hand side of the above inequality is written as
    \[
         \frac{D^{a^R_{j_0}}}{D^{a^R_j}}{\tilde{\lambda}_0^{a_{j_0}^L}}=\frac{K_j}{L_j}.
    \]
    Therefore, we obtain
    \[
        \frac{\lambda_0^{a_{j}^L}}{K_j}\leqq \frac{1}{L_j}.
    \]
    For each $j\in\left\{0,\ldots,J-1\right\}$, the majorization condition for LOCC convertibility between bipartite pure states~\cite{N2} guarantees that there exists an LOCC map represented by a family of operators ${\left\{M_{j,m_1}\otimes U_{j,m_1}\right\}}_{m_1}$ achieving, for each $m_1$,
    \begin{align*}
        \left(M_{j,m_1}\otimes U_{j,m_1}\right)\left(\Ket{\omega_j}^{a^L b^L}\otimes\Ket{\Phi^+_{K_j}}^{\overline{A}\overline{B}}\right)
        =\Ket{\Phi^+_{L_j}}^{\overline{A}\overline{B}},
    \end{align*}
    where ${\left\{M_{j,m_1}\right\}}_{m_1}$ represents $A$'s measurement from $\mathcal{H}^{a^L}\otimes\mathcal{H}^{\overline{A}}$ to $\mathcal{H}^{\overline{A}}$ with outcome $m_1$ satisfying the completeness $\sum_{m_1} M_{j,m_1}^\dag M_{j,m_1}=\mathbbm{1}$, and $U_{j,m_1}$ represents $B$'s isometry from $\mathcal{H}^{b^L}\otimes\mathcal{H}^{\overline{B}}$ to $\mathcal{H}^{\overline{B}}$ conditioned by $m_1$.
    Regarding an explicit form of ${\left\{M_{j,m_1}\otimes U_{j,m_1}\right\}}_{m_1}$, refer to Refs.~\cite{N2,T3}.

    \textit{Subprocess~2: Quantum teleportation to transfer the quantum part.}
    While quantum teleportation for sending the full reduced state $\phi_j^{a^R}\coloneqq\tr_{R b^R}\Ket{\phi_j}\Bra{\phi_j}^{R a^R b^R}$ requires a maximally entangled resource state with Schmidt rank
    \[
      \dim\mathcal{H}^{a^R}=\max_j\dim\mathcal{H}^{a_j^R},
    \]
    we adopt a compression method instead of just performing quantum teleportation of $\phi_j^{a^R}$, so that each $\phi_j^{a^R}$ is transferred from $A$ to $B$ using a maximally entangled resource state with Schmidt rank $\dim\mathcal{H}^{a_j^R}$, which is smaller than or equal to $\dim\mathcal{H}^{a^R}$.
    Consider $A$'s auxiliary system $\bigotimes_{j=0}^{J-1}\mathcal{H}^{{\left(a'\right)}_j^R}$, where $\dim\mathcal{H}^{{\left(a'\right)}_j^R}=D^{a_j^R}$.
    In our algorithm, $\Ket{\phi_j}^{R a^R b^R}$ is compressed into
    \[
        \begin{split}
            \Ket{\phi_j}^{R{\left(a'\right)}_j^R b^R}=U'_j\Ket{\phi_j}^{R a^R b^R},
        \end{split}
    \]
    where $U'_j$ is an isometry from $\mathcal{H}^{a^R}$ to $\mathcal{H}^{{\left(a'\right)}_j^R}$, and $\Ket{\phi_j}^{R{\left(a'\right)}_j^R b^R}$ represents the same state as $\Ket{\phi_j}^{R a^R b^R}$.
    Quantum teleportation~\cite{B5} to send states of $\mathcal{H}^{{\left(a'\right)}_j^R}$ consists of $A$'s projective measurement in the maximally entangled basis ${\left\{\Ket{\Phi_{j,m_2}}\right\}}_{m_2}$ on $\mathcal{H}^{{\left(a'\right)}_j^R}\otimes\mathcal{H}^{\overline{A}}$ with outcome $m_2$ and $B$'s generalized Pauli correction $\sigma_{j,m_2}$ from $\mathcal{H}^{\overline{B}}$ to $\mathcal{H}^{{(b')}^R}$ conditioned by $m_2$, where $\mathcal{H}^{{(b')}^R}$ is $B$'s auxiliary system corresponding to $\mathcal{H}^{a^R}$.
    The map for quantum teleportation is represented by ${\left\{\Bra{\Phi_{j,m_2}}\otimes\sigma_{j,m_2}\right\}}_{m_2}$, which traces out the post-measurement state of $A$ and achieves, for each $m_2$,
    \begin{equation*}
        \begin{split}
            &\left(\Bra{\Phi_{j,m_2}}\otimes\sigma_{j,m_2}\right)\left(\Ket{\phi_j}^{R {\left(a'\right)}_j^R b^R}\otimes\Ket{\Phi_{D^{a_j^R}}^+}^{\overline{A}\overline{B}}\right)\\
            &=\left[\left(\Bra{\Phi_{j,m_2}}U'_j\right)\otimes\sigma_{j,m_2}\right]\left(\Ket{\phi_j}^{R a^R b^R}\otimes\Ket{\Phi_{D^{a_j^R}}^+}^{\overline{A}\overline{B}}\right)\\
            &=\Ket{\phi_j}^{R {(b')}^R b^R}.
        \end{split}
    \end{equation*}

    \textit{Subprocess~3: Coherently merging the classical part by a measurement.}
    As for the classical part $\mathcal{H}^{a_0}$,
    a measurement should be performed by $A$ to merge the classical part without breaking coherence between $B$ and $R$.
    This contrasts with the algorithm proposed in Ref.~\cite{K4} for transferring a state drawn from a given ensemble, in which a projective measurement onto each of the subspaces of the Koashi-Imoto decomposition indexed by $j$ destroys superposition of states among different subspaces.
    In our algorithm, $A$'s measurement on $\mathcal{H}^{a_0}$ is a projective measurement with outcome $m_3$ in the Fourier basis ${\left\{\Ket{m_3}\right\}}_{m_3}$ defined in terms of the computational basis ${\left\{\Ket{j}^{a_0}\right\}}_j$, that is, for each $m_3$,
    \begin{equation*}
        \Ket{m_3}^{a_0}\coloneqq \sum_{j=0}^{J-1}\exp\left(\frac{\textup{i}{\pi}jm_3}{J}\right)\Ket{j}^{a_0}.
    \end{equation*}
    After sending the measurement outcome $m_3$ by classical communication from $A$ to $B$,
    the originally given state of $\mathcal{H}^{a_0}\otimes\mathcal{H}^{a_L}\otimes\mathcal{H}^{b_L}$ can be recovered from $B$'s classical part $\mathcal{H}^{b_0}$ of the post-measurement state by $B$'s local isometry conditioned by $m_3$
    \begin{equation}
      \label{eq:subprocess_3}
      \sum_{j=0}^{J-1}\exp\left(\frac{\textup{i}{\pi}jm_3}{J}\right)\Ket{j}^{{(b')}_0}\otimes\Ket{j}\Bra{j}^{b_0}\otimes\Ket{\omega_j}^{{(b')}^L b^L},
    \end{equation}
    where $\mathcal{H}^{{(b')}_0}\otimes\mathcal{H}^{{(b')}^L}$ is $B$'s auxiliary system corresponding to $\mathcal{H}^{a_0}\otimes\mathcal{H}^{a^L}$.

    We combine Subprocesses~1--3 using controlled measurements and controlled isometries.
    Regarding $A$'s measurement,
    the measurements used in Subprocesses~1 and~2 are performed by extending each measurement to a measurement controlled coherently by the computational-basis state $\Ket{j}^{a_0}$.
    Regarding Subprocess~1 for the redundant part, the controlled version of the measurement is given by
    $\sum_{j=0}^{J-1}\Ket{j}\Bra{j}^{a_0}\otimes M_{j,m_1}$,
    and regarding Subprocess~2 for the quantum part, given by
    $\sum_{j=0}^{J-1}\Ket{j}\Bra{j}^{a_0}\otimes\left(\Bra{\Phi_{j,m_2}}U'_j\right)$.
    The measurement in Subprocess~3 for the classical part is also represented in terms of the computational basis as
    $\sum_{j=0}^{J-1}\Bra{m_3}^{a_0}\left(\Ket{j}\Bra{j}^{a_0}\right)=\sum_{j=0}^{J-1}\exp\left(\frac{-\textup{i}{\pi}jm_3}{J}\right)\Bra{j}^{a_0}$.
    Combining these three together, we obtain $A$'s measurement ${\left\{M_{m_1,m_2,m_3}\right\}}_{m_1,m_2,m_3}$ given by
    \[
        \begin{split}
          &M_{m_1,m_2,m_3}\\
          &=\sum_{j=0}^{J-1}\left[\exp\left(\frac{-\textup{i}{\pi}jm_3}{J}\right)\Bra{j}^{a_0}\right]\otimes\left[\Bra{\Phi_{j,m_2}}U'_j M_{j,m_1}\right].
        \end{split}
    \]
    The completeness of this measurement follows from
    \[
        \begin{split}
            &\sum_{m_1,m_2,m_3}M_{m_1,m_2,m_3}^\dag M_{m_1,m_2,m_3}\\
            &=\sum_{m_1,m_2,m_3}\sum_{j,j'}\left[\exp\left(\frac{\textup{i}{\pi}m_3(j'-j)}{J}\right)\Ket{j'}\Bra{j}\right]\\
            &\quad\otimes\left[M_{j,m_1}^\dag {U'_{j'}}^\dag\Ket{\Phi_{j',m_2}}\Bra{\Phi_{j,m_2}}U'_j M_{j,m_1}\right]\\
            &=\sum_{j}\Ket{j}\Bra{j}
            \otimes\left[\sum_{m_1,m_2}M_{j,m_1}^\dag{U'_j}^\dag\Ket{\Phi_{j,m_2}}\Bra{\Phi_{j,m_2}}U'_j M_{j,m_1}\right]\\
            &=\mathbbm{1},
        \end{split}
    \]
    where $\mathbbm{1}$ is the identity operator on $\mathcal{H}^{a_0}\otimes\mathcal{H}^{a^L}\otimes\mathcal{H}^{a^R}\otimes\mathcal{H}^{\overline{A}}$.

    As for $B$'s isometry,
    the isometries in Subprocesses~1 and~2 are also controlled coherently by the computational-basis state $\Ket{j}^{b_0}$.
    Regarding Subprocess~1 for the redundant part, the controlled version of the isometry is given by
    $\sum_{j=0}^{J-1}\Ket{j}\Bra{j}^{b_0}\otimes U_{j,m_1}$,
    and regarding Subprocess~2 for the quantum part, given by
    $\sum_{j=0}^{J-1}\Ket{j}\Bra{j}^{b_0}\otimes\sigma_{j,m_2}$.
    The isometry in Subprocess~3 is given by Eq.~\eqref{eq:subprocess_3}.
    Combining these three together, we obtain $B$'s isometry $U_{m_1,m_2,m_3}$ given by
    \[
        \begin{split}
          &U_{m_1,m_2,m_3}\\
          &=\sum_{j=0}^{J-1}\exp\left(\frac{\textup{i}{\pi}jm_3}{J}\right)\Ket{j}^{{(b')}_0}\otimes\Ket{j}\Bra{j}^{b_0}\otimes\Ket{\omega_j}^{{(b')}^L b^L}\\
          &\quad\otimes\sigma_{j,m_2} U_{j,m_1}.
        \end{split}
    \]

    Consequently, for any combination $(m_1,m_2,m_3)$,
    the LOCC map represented by a family of operators
    \[
      {\left\{M_{m_1,m_2,m_3}\otimes U_{m_1,m_2,m_3}\right\}}_{m_1,m_2,m_3}
    \]
    acts as
    \begin{equation}
      \label{eq:transformation_merging}
      \begin{split}
        &\left(M_{m_1,m_2,m_3}\otimes U_{m_1,m_2,m_3}\right)\\
        &\quad\left(\Ket{j}^{a_0}\otimes\Ket{j}^{b_0}\otimes\Ket{\omega_j}^{a^L b^L}\otimes\Ket{\phi_j}^{R a^R b^R}\right.\\
        &\qquad\left.\otimes\Ket{\Phi_{D^{a_j^R}}^+}^{\overline{A}\overline{B}}\otimes\Ket{\Phi_{K_j}^+}^{\overline{A}\overline{B}}\right)\\
        &=\Ket{j}^{{(b')}_0}\otimes\Ket{j}^{b_0}\otimes\Ket{\omega_j}^{{(b')}^L b^L}\otimes\Ket{\phi_j}^{R {(b')}^R b^R}
        \otimes\Ket{\Phi_{L_j}^+}^{\overline{A}\overline{B}}.
      \end{split}
    \end{equation}
    For each $j$, the entanglement cost is evaluated by
    \begin{align*}
        &\log_2 D^{a_j^R} + \log_2 K_j - \log_2 L_j\\
        &= \log_2 \left(\frac{K_j}{L_j} D^{a_j^R}\right)\\
        &= \log_2 \left(\tilde{\lambda}_0^{a_{j_0}^L}D^{a_{j_0}^R}\right),
    \end{align*}
    which is independent of $j$.
    Choosing $K$ as the least common multiple of the integers $\left\{D^{a_0^R}K_0,\ldots,D^{a_{J-1}^R}K_{J-1}\right\}$,
    we can rewrite Eq.~\eqref{eq:transformation_merging} as
    \[
      \begin{split}
        &\left(M_{m_1,m_2,m_3}\otimes U_{m_1,m_2,m_3}\right)\\
        &\quad\left(\Ket{j}^{a_0}\otimes\Ket{j}^{b_0}\otimes\Ket{\omega_j}^{a^L b^L}\otimes\Ket{\phi_j}^{R a^R b^R}\otimes\Ket{\Phi_{K}^+}^{\overline{A}\overline{B}}\right)\\
        &=\Ket{j}^{{(b')}_0}\otimes\Ket{j}^{b_0}\otimes\Ket{\omega_j}^{{(b')}^L b^L}\otimes\Ket{\phi_j}^{R {(b')}^R b^R}\otimes\Ket{\Phi_{L}^+}^{\overline{A}\overline{B}},
      \end{split}
    \]
    where $L$ is an integer defined as
    \[
      L\coloneqq\frac{K}{\tilde{\lambda}_0^{a_{j_0}^L}D^{a_{j_0}^R}}=\frac{K}{D^{a_j^R} K_j}L_j,\quad \forall j\in\left\{0,\ldots,J-1\right\}.
    \]
    Thus, we obtain an LOCC map represented by
    \begin{equation}
      \label{eq:merge}
      \begin{split}
        &\left\{\left[ M_{m_1,m_2,m_3}U^A\right]\right.\\
        &{\quad\left.\otimes\left[
    \left({\left(U^{B'}\right)}^\dag\otimes {\left(U^B\right)}^\dag\right) U_{m_1,m_2,m_3}U^B\right]\right\}}_{m_1,m_2,m_3},
      \end{split}
    \end{equation}
    which achieves for each $(m_1,m_2,m_3)$
    \begin{equation*}
        \begin{split}
          &\left[ M_{m_1,m_2,m_3}U^A\right]\otimes\left[
        \left({\left(U^{B'}\right)}^\dag\otimes {\left(U^B\right)}^\dag\right) U_{m_1,m_2,m_3}U^B\right]\\
        &\quad\Ket{\psi}^{RAB}\otimes\Ket{\Phi_{K}^+}^{\overline{A}\overline{B}}\\
        &=\Ket{\psi}^{RB'B}\otimes\Ket{\Phi_{L}^+}^{\overline{A}\overline{B}},
      \end{split}
    \end{equation*}
    where $U^A$ and $U^B$ are those in Eq.~\eqref{eq:ki_tripartite_isometry}, and ${\left(U^{B'}\right)}^\dag$ from $\mathcal{H}^{{(b')}_0}\otimes\mathcal{H}^{{(b')}^L}\otimes\mathcal{H}^{{(b')}^R}$ to $\mathcal{H}^{B'}=\bigoplus_{j=0}^{J-1}\mathcal{H}^{{(b')}_j^L}\otimes\mathcal{H}^{{(b')}_j^R}$ acts in the same way as ${\left(U^A\right)}^\dag$.
    The entanglement cost of the algorithm represented by the LOCC map shown in Eq.~\eqref{eq:merge} is given by
    \begin{align*}
      &\log_2 K-\log_2 L\\
      &=\log_2 \left(\tilde{\lambda}_0^{a_{j_0}^L}D^{a_{j_0}^R}\right)\\
      &\leqq \log_2 \left(\lambda_0^{a_{j_0}^L}D^{a^R_{j_0}}\right)+\delta\\
      &=\max_j\left\{\log_2 \left(\lambda_0^{a_{j_0}^L}\dim\mathcal{H}^{a^R_{j_0}}\right)\right\}+\delta,
    \end{align*}
    which yields the conclusion.
\end{IEEEproof}

\begin{IEEEproof}[Proof of Theorem~\ref{thm:merge_without_catalyst}]
    We construct an algorithm for non-catalytic exact state merging of $\Ket{\psi}^{RAB}$ achieving the equality in~\eqref{eq:merge_without_catalyst_cost}.
    We define, for each $j\in\{0,\ldots,J-1\}$,
    \begin{align*}
        D^{a_j^R}&\coloneqq\dim\mathcal{H}^{a_j^R}.
    \end{align*}
    We omit identity operators, such as $\mathbbm{1}^R$, in the following for brevity.
    The core idea of the algorithm is similar to that in Theorem~\ref{thm:merge} using the Koashi-Imoto decomposition in the form of Eq.~\eqref{eq:ki_tripartite_isometry}.
    The rest of the proof is given in the same way as the proof of Theorem~\ref{thm:merge}, where Subprocess~2 and Subprocess~3 are the same as those in Theorem~\ref{thm:merge}, and Subprocess~1 is modified as follows since we do not use the resource state catalytically in the entanglement distillation from the redundant part in Subprocess~1.

    \textit{Subprocess~1:}
    For each $j\in\{0,\ldots,J-1\}$, it holds that
    \begin{align*}
        \lambda_{0}^{a_{j}^L}D^{a^R_j}
        \leqq\left\lceil\lambda_{0}^{a_j^L}D^{a_j^R}\right\rceil
        \leqq\max_j\left\{ \left\lceil \lambda_0^{a_j^L}D^{a^R_j}\right\rceil\right\}.
    \end{align*}
    Then, given the resource state $\Ket{\Phi_K^+}$, where
    \[
        K=\max_j\left\{\left\lceil \lambda_0^{a_j^L}D^{a_j^R}\right\rceil\right\},
    \]
    we have
    \[
        \frac{\lambda_{0}^{a_{j}^L}}{K}\leqq\frac{1}{D^{a_j^R}}.
    \]
    For each $j\in\left\{0,\ldots,J-1\right\}$, the majorization condition for LOCC convertibility between bipartite pure states~\cite{N2} guarantees that there exists an LOCC map represented by a family of operators ${\left\{M_{j,m_1}\otimes U_{j,m_1}\right\}}_{m_1}$ achieving, for each $m_1$,
    \begin{align*}
        \left(M_{j,m_1}\otimes U_{j,m_1}\right)\left(\Ket{\omega_j}^{a^L b^L}\otimes\Ket{\Phi^+_{K}}^{\overline{A}\overline{B}}\right)
        =\Ket{\Phi^+_{D^{a_j^R}}}^{\overline{A}\overline{B}},
    \end{align*}
    where ${\left\{M_{j,m_1}\right\}}_{m_1}$ represents $A$'s measurement from $\mathcal{H}^{a^L}\otimes\mathcal{H}^{\overline{A}}$ to $\mathcal{H}^{\overline{A}}$ with outcome $m_1$ satisfying the completeness $\sum_{m_1} M_{j,m_1}^\dag M_{j,m_1}=\mathbbm{1}$, and $U_{j,m_1}$ represents $B$'s isometry from $\mathcal{H}^{b^L}\otimes\mathcal{H}^{\overline{B}}$ to $\mathcal{H}^{\overline{B}}$ conditioned by $m_1$.

    In the same way as Theorem~\ref{thm:merge},
    $A$'s combined measurement ${\left\{\Bra{m_1,m_2,m_3}\right\}}_{m_1,m_2,m_3}$, where the post-measurement state is traced out, is given by
    \[
        \begin{split}
          &\Bra{m_1,m_2,m_3}\\
          &=\sum_{j=0}^{J-1}\left[\exp\left(\frac{-\textup{i}{\pi}jm_3}{J}\right)\Bra{j}^{a_0}\right]\otimes\left[\Bra{\Phi_{j,m_2}}U'_j M_{j,m_1}\right].
        \end{split}
    \]

    Also, $B$'s combined isometry $U_{m_1,m_2,m_3}$ is given by
    \[
        \begin{split}
          &U_{m_1,m_2,m_3}\\
          &=\sum_{j=0}^{J-1}\exp\left(\frac{\textup{i}{\pi}jm_3}{J}\right)\Ket{j}^{{(b')}_0}\otimes\Ket{j}\Bra{j}^{b_0}\otimes\Ket{\omega_j}^{{(b')}^L b^L}\\
          &\quad\otimes\sigma_{j,m_2} U_{j,m_1}.
        \end{split}
    \]

    Consequently, we obtain an LOCC map represented by
    \begin{equation}
      \label{eq:merge_without_catalyst}
      \begin{split}
        &\left\{\left[\Bra{m_1,m_2,m_3}U^A\right]\right.\\
        &{\quad\left.\otimes
    \left[\left({\left(U^{B'}\right)}^\dag\otimes{\left(U^B\right)}^\dag\right) U_{m_1,m_2,m_3}U^B\right]\right\}}_{m_1,m_2,m_3},
      \end{split}
    \end{equation}
    which achieves, for any combination $(m_1,m_2,m_3)$,
    \begin{equation*}
      \begin{split}
        &\left[\Bra{m_1,m_2,m_3}U^A\right]\otimes
        \left[\left({\left(U^{B'}\right)}^\dag\otimes{\left(U^B\right)}^\dag\right) U_{m_1,m_2,m_3}U^B\right]\\
        &\quad\Ket{\psi}^{RAB}\otimes\Ket{\Phi_{K}^+}^{\overline{A}\overline{B}}\\
        &=\Ket{\psi}^{RB'B},
      \end{split}
    \end{equation*}
    where $U^A$, $U^B$, and $U^{B'}$ are the same as those in Eq.~\eqref{eq:merge}.
    The entanglement cost of the algorithm represented by the LOCC map shown in Eq.~\eqref{eq:merge_without_catalyst} is given by
    \[
      \begin{split}
        &\log_2 K\\
        &= \max_j\left\{\log_2\left\lceil \lambda_0^{a_j^L}D^{a_j^R}\right\rceil\right\}\\
        &=\log_2\max_j\left\{\log_2\left\lceil \lambda_0^{a_j^L}\dim\mathcal{H}^{a_j^R}\right\rceil\right\},
      \end{split}
    \]
    which yields the conclusion.
\end{IEEEproof}

\begin{remark}
    \textit{Comparison between exact state merging and splitting.}
    Entanglement cost in exact state merging is not larger than that in its inverse task, that is, exact state splitting analyzed in Appendix~\ref{sec:split}.
    For any $\Ket{\psi}^{RAB}$,
    \[
        \begin{split}
          &\max_{j}\left\{\log_2\lambda_0^{a_j^L}\dim\mathcal{H}^{a_j^R}\right\}\leqq\log_2\rank\psi^{A},\\
          &\max_{j}\left\{\log_2\left\lceil\lambda_0^{a_j^L}\dim\mathcal{H}^{a_j^R}\right\rceil\right\}\leqq\log_2\rank\psi^{A},
        \end{split}
    \]
    where the right-hand sides are the optimal entanglement cost in exact state splitting obtained in Theorem~\ref{thm:split} in Appendix~\ref{sec:split_result}, and the notations are the same as those in Theorems~\ref{thm:merge} and~\ref{thm:merge_without_catalyst}.
    These inequalities can be derived from $\dim\mathcal{H}^{a_j^R}\leqq\rank\psi^{A}$ and $\lambda_0^{a_j^L}\leqq 1$,
    where the former inequality holds by construction of the Koashi-Imoto decomposition.
    Moreover, as shown in Implication~\ref{ex:1} in Sec.~\ref{sec:examples}, entanglement cost in exact state merging can be strictly smaller than that in spitting.
\end{remark}

\begin{remark}
\label{remark:usefulness}
    \textit{Usefulness of the algorithms for exact state merging on small and intermediate scales.}
    We discuss the cases where the obtained algorithms for exact state merging outperforms the existing algorithms for one-shot approximate state merging~\cite{B9,Y9,B12,D7,D6,H10,B10,D5,M,N3,A4,A5} in terms of entanglement cost.

    For a given approximation error $\epsilon >0$,
    the algorithms for one-shot approximate state merging of $\Ket{\psi}^{RAB}$ transform ${\psi}^{RAB}\otimes{\Phi_K^+}^{\overline{A}\overline{B}}$ into a final state $\psi_\textup{final}$ satisfying ${F^2\left(\psi^{RB^\prime B}\otimes{\Phi_L^+}^{\overline{A}\overline{B}},\psi_\textup{final}\right)}\coloneqq\left(\Bra{\psi}\otimes\Bra{\Phi_L^+}\right)\psi_\textup{final}\left(\Ket{\psi}\otimes\Ket{\Phi_L^+}\right)\geqq 1-\epsilon^2$, where $F$ represents the fidelity.
    While some of the existing algorithms are fully quantum algorithms achieved by local operations and quantum communication assisted by shared entanglement,
    we replace the quantum communication in a fully quantum algorithm with quantum teleportation to obtain an entanglement-assisted LOCC algorithm corresponding to the fully quantum algorithm and compare entanglement cost $E(\psi)$ in one-shot state merging of $\Ket{\psi}^{RAB}$ in the LOCC framework.

    Our algorithms for exact state merging of $\Ket{\psi}^{RAB}$ require at most as much entanglement cost as quantum teleportation of $\psi^A$, and when the system size for $\psi^A$ is small, our algorithms cost less than the existing algorithms for one-shot approximate state merging.
    Regarding the existing algorithms,
    the achievability bounds of $E(\psi)$ of the corresponding entanglement-assisted LOCC algorithms can be calculated from the analyses in Refs.~\cite{B9,B12,D7,D6,H10,B10,N3}.
    Given $\epsilon > 0$, these achievability bounds are in the form
    $E(\psi)=\cdots+O\left(\log\frac{1}{\epsilon}\right)$
    as $\epsilon\to 0$, which diverges to infinity as higher fidelity is pursued.
    For example, from Theorem~4 in Ref.~\cite{B10}, the achievability bound of $E(\psi)$ of one-shot state merging of $\Ket{\psi}^{RAB}$ within an error $\epsilon>0$ is given by
    \[
      {H_{\max}^{\epsilon_1}\left(A|B\right)}_{\psi}+2\log_2 \frac{1}{\epsilon_4}+3,
    \]
    where $\epsilon=8\epsilon_1+\sqrt{3\epsilon_4}$, and the first term is represented by the smooth conditional max-entropy~\cite{R2,T5}.
    To achieve $\epsilon=0.02$, the second and third terms amount to
    \[
      2\log_2 \frac{1}{\epsilon_4}+3>28.7.
    \]
    Note that $\epsilon=0.02$ guarantees, in the task of state discrimination of $\Ket{\psi}$ and $\psi_\textup{final}$, the optimal success probability $P_\textup{succ}=\frac{1}{2}+\frac{1}{4}{\left\|\psi-\psi_\textup{final}\right\|}_1\leqq 51\%$, which is obtained from the Fuchs-van de Graaf inequalities $\frac{1}{4}{\left\|\psi-\psi_\textup{final}\right\|}_1\leqq \frac{1}{2}\sqrt{1-F^2}$~\cite{W11}.
    Thus, given $\Ket{\psi}^{RAB}$ where $\dim\mathcal{H}^A\leqq 2^{28}$, even if ${H_{\max}^{\epsilon_1}\left(A|B\right)}_{\psi}=0$, the approximate algorithm requires more entanglement cost than our algorithms and even than quantum teleportation.
\end{remark}

\begin{remark}
\label{remark:approximate}
  \textit{Extension of our algorithms to approximate state merging.}
  Our algorithms for exact state merging can be extended to approximate state merging by means of smoothing~\cite{R2,T5}, as presented in Appendix~\ref{sec:approximate}.
  The achievability bound of entanglement cost in approximate state merging of a given state $\Ket{\psi}^{RAB}$ within a given error $\epsilon\geqq 0$ is shown in Theorem~\ref{thm:approximate} in Appendix~\ref{sec:approximate}.
  Note that this achievability bound includes minimization over any state which is $\frac{\epsilon}{2}$-close to $\Ket{\psi}^{RAB}$ in terms of fidelity,
  and no simple strategy is known to evaluate this minimization in general as the direct-sum structure of the Koashi-Imoto decomposition may discontinuously change under smoothing.
  However, as will be discussed in Implication~\ref{ex:1} in Sec.~\ref{sec:examples}, useful states for distributed quantum information processing, including the Greenberger-Horne-Zeilinger (GHZ) states and multipartite code states for quantum error correcting codes, have \textit{nontrivial} Koashi-Imoto decomposition, that is, $J\neq 1$, when these states are regarded as tripartite states.
  In this regard, the algorithms for exact state merging are already sufficient for reducing entanglement cost compared to quantum teleportation in these cases relevant to distributed quantum information processing.
\end{remark}

\subsection{\label{sec:converse}Improved converse bound for exact state merging}

We provide a converse bound of entanglement cost of exact state merging.
This converse bound improves the existing converse bound in terms of conditional max-entropy originally shown in Ref.~\cite{B9}.
In this section, after showing our bound, we compare the bound with the existing bound and then discuss the tightness of the bound.

Our converse bound for exact state merging is shown as follows.

\begin{theorem}
\label{thm:new}
\textit{A converse bound of entanglement cost in exact state merging.}
For any state $\Ket{\psi}^{RAB}$ and any algorithm for exact state merging of $\Ket{\psi}^{RAB}$,
it holds that
\begin{equation}
    \label{eq:lower_catalytic}
    \begin{split}
      &\log_2 K - \log_2 L\\
      &\geqq \inf\left\{\log_2 K - \log_2 L: \frac{\mathbbm{1}_K}{K}\otimes\psi^{B}\prec\frac{\mathbbm{1}_L}{L}\otimes\psi^{AB}\right\},
    \end{split}
\end{equation}
where $\prec$ denotes majorization for hermitian operators~\cite{W11}.
Also, for any algorithm for non-catalytic exact state merging of $\Ket{\psi}^{RAB}$,
it holds that
\begin{equation}
    \label{eq:lower_non_catalytic}
    \begin{split}
      &\log_2 K\\
      &\geqq \min\left\{\log_2 K: \frac{\mathbbm{1}_K}{K}\otimes\psi^{B}\prec\psi^{AB}\right\},
    \end{split}
\end{equation}
where the notations are the same as those in Inequality~\eqref{eq:lower_catalytic}.
\end{theorem}

\begin{IEEEproof}
  We prove Inequality~\eqref{eq:lower_catalytic}, while Inequality~\eqref{eq:lower_non_catalytic} can be shown in a similar way by substituting $L$ in the following proof with $1$.

  Any algorithm for exact state merging transforms $\Ket{\psi}^{RAB}\otimes\Ket{\Phi_K^+}^{\overline{A}\overline{B}}$ into $\Ket{\psi}^{RB'B}\otimes\Ket{\Phi_L^+}^{\overline{A}\overline{B}}$ by LOCC\@.
  Hence, with respect to the bipartition between $\mathcal{H}^R\otimes\mathcal{H}^A\otimes\mathcal{H}^{\overline{A}}$ and $\mathcal{H}^B\otimes\mathcal{H}^{B^\prime}\otimes\mathcal{H}^{\overline{B}}$, LOCC convertibility between bipartite pure states yields
  the majorization condition~\cite{N2}
  \[
    \frac{\mathbbm{1}_K}{K}\otimes\psi^{B}\prec\frac{\mathbbm{1}_L}{L}\otimes\psi^{AB}
  \]
  in terms of hermitian operators representing their reduced states.
  Since this majorization holds for any $K$ and $L$ achieving exact state merging of $\Ket{\psi}^{RAB}$, we obtain Inequality~\eqref{eq:lower_catalytic}.
\end{IEEEproof}

As a corollary of Theorem~\ref{thm:new}, we obtain the following converse bound for states in the form of $\Ket{\Phi_D^+\left(\psi\right)}^{RAB}$ defined as Eq.~\eqref{eq:max}, which is easier to calculate than that in Theorem~\ref{thm:new}.
The following analysis in this section may assume that $\psi^R=\frac{\mathbbm{1}^R}{D}$ holds for a given state $\Ket{\psi}^{RAB}$ for simplicity,
based on the fact that entanglement cost in exact state merging of $\Ket{\psi}^{RAB}$ and that of $\Ket{\Phi_D^+\left(\psi\right)}^{RAB}$ are the same, as discussed in Sec.~\ref{sec:def_merge} as well as Appendix~\ref{sec:equivalence}.
Note that to calculate the converse bound in the following corollary for any given state $\Ket{\psi}^{RAB}$, first calculate the Schmidt decomposition of $\Ket{\psi}^{RAB}$ to obtain the corresponding maximally entangled state $\Ket{\Phi_D^+\left(\psi\right)}^{RAB}$ from Eq.~\eqref{eq:max}, and then apply the corollary.

\begin{corollary}
\label{col:tractable_converse}
  \textit{A converse bound of entanglement cost in exact state merging derived from Theorem~\ref{thm:new}}.
  For any state $\Ket{\psi}^{RAB}$ satisfying $\psi^R=\frac{\mathbbm{1}^R}{D}$, and any algorithm for exact state merging of $\Ket{\psi}^{RAB}$, it holds that
  \begin{equation}
    \label{eq:tractable_lower_catalytic}
    \log_2 K - \log_2 L \geqq \log_2 \left({\lambda_0^B}D\right),
  \end{equation}
  where $\lambda_0^B$ is the largest eigenvalue of $\psi^B$.
  Also, for any algorithm for non-catalytic exact state merging of $\Ket{\psi}^{RAB}$ satisfying $\psi^R=\frac{\mathbbm{1}^R}{D}$,
  it holds that
  \begin{equation}
    \label{eq:tractable_lower_non_catalytic}
    \log_2 K \geqq \log_2 \left\lceil\lambda_0^B D\right\rceil,
  \end{equation}
  where $\lceil{}\cdots{}\rceil$ is the ceiling function, and $\lambda_0^B$ is the same as that in Eq.~\eqref{eq:tractable_lower_catalytic}.
\end{corollary}

\begin{IEEEproof}[\textit{Proof of Inequality~\eqref{eq:tractable_lower_catalytic}}]
  Due to Theorem~\ref{thm:new}, exact state merging implies
  \[
    \frac{\mathbbm{1}_K}{K}\otimes\psi^{B}\prec\frac{\mathbbm{1}_L}{L}\otimes\psi^{AB}.
  \]
  Thus, the largest eigenvalues of the both sides of this majorization satisfy
  \[
    \frac{\lambda_0^B}{K}\leqq\frac{1}{DL},
  \]
  and we obtain
  \[
    \log_2 K - \log_2 L \geqq \log_2\left({\lambda_0^B}D\right).
  \]
\end{IEEEproof}

\begin{IEEEproof}[\textit{Proof of Inequality~\eqref{eq:tractable_lower_non_catalytic}}]
    From the same argument as the above, we obtain
    \[
        \frac{\lambda_0^B}{K}\leqq\frac{1}{D}.
    \]
    Hence, it holds that
    \[
        K\geqq\lambda_0^B D,
    \]
    and since $K$ is an integer, we have
    \[
        K\geqq \left\lceil \lambda_0^B D\right\rceil.
    \]
    Therefore, we obtain
    \[
        \log_2 K \geqq \log_2 \left\lceil\lambda_0^B D\right\rceil.
    \]
\end{IEEEproof}

Reference~\cite{B9} also provides a converse bound of entanglement cost in exact state merging of any given state $\Ket{\psi}^{RAB}$ in terms of the conditional max-entropy as follows.
Note that this converse bound in Ref.~\cite{B9} is only shown for one-way LOCC, while our converse bounds in Theorem~\ref{thm:new} and Corollary~\ref{col:tractable_converse} are applicable to any LOCC map including two-way LOCC\@.

\begin{lemma}
\label{lem:old}
    (Corollary 4.12.\ in Ref.~\cite{B9})
    \textit{A converse bound of entanglement cost in exact state merging in Ref.~\cite{B9}.}
    For any state $\Ket{\psi}^{RAB}$ and any one-way LOCC algorithm for exact state merging of $\Ket{\psi}^{RAB}$,
    where classical communication is performed only from $A$ to $B$,
    it holds that
    \begin{equation*}
        \log_2 K - \log_2 L \geqq {H_{\max}(A|B)}_\psi,
    \end{equation*}
    where the right-hand side is the conditional max-entropy~\cite{R2,T5}.
\end{lemma}

For states in the form of Eq.~\eqref{eq:max}, our converse bounds in Theorem~\ref{thm:new} and Corollary~\ref{col:tractable_converse} are at least as tight as the existing bound in Lemma~\ref{lem:old}, as shown in the following proposition.
Moreover, Implication~\ref{ex:2} in Sec.~\ref{sec:examples} will show a case where our bound is strictly tighter than the existing bound.
Note that while Corollary~\ref{col:tractable_converse} assumes states in the form of Eq.~\eqref{eq:max}, the converse bounds in Theorem~\ref{thm:new} and Lemma~\ref{lem:old} also hold without this assumption.
It is sufficient to show that the converse bound in Corollary~\ref{col:tractable_converse} is at least as tight as that in Lemma~\ref{lem:old}, since Theorem~\ref{thm:new} provides at least as tight bound as that in Corollary~\ref{col:tractable_converse}.

\begin{proposition}
  \textit{Comparison of converse bounds of entanglement cost in exact state merging.}
    For any state $\Ket{\psi}^{RAB}$ satisfying $\psi^R=\frac{\mathbbm{1}^R}{D}$,
    it holds that
    \[
        \log_2\left({\lambda_0^B}D\right)\geqq{H_{\max}(A|B)}_\psi,
    \]
    where the notations are the same as those in Corollary~\ref{col:tractable_converse} and Lemma~\ref{lem:old}.
\end{proposition}

\begin{IEEEproof}
    We write the Schmidt decomposition of $\Ket{\psi}^{RAB}$ as
    \[
        \Ket{\psi}^{RAB}=\sum_{l=0}^{D-1} \frac{1}{\sqrt{D}}\Ket{l}^R\otimes\Ket{\psi_l}^{AB}.
    \]
    Reference~\cite{V2} provides a semidefinite programming for $2^{{H_{\max}(A|B)}_\psi}$:
    minimize ${\left\|Z^B\right\|}_\infty$ subject to $\mathbbm{1}^R \otimes Z^{AB}\geqq\Ket{\psi}\Bra{\psi}^{RAB}$ and $Z^{AB}\geqq 0$.
    The case $Z^{AB}=D\psi^{AB}$ satisfies these constraints:
    \[
        \begin{split}
            &\mathbbm{1}^R\otimes D\psi^{AB}=\sum_l\Ket{l}\Bra{l}^R\otimes\sum_l\Ket{\psi_l}\Bra{\psi_l}^{AB}\geqq\Ket{\psi}\Bra{\psi}^{RAB};\\
            &D\psi^{AB}\geqq 0.
        \end{split}
    \]
    Therefore,
    \[
      \begin{split}
        &\log_2 \left({\lambda_0^B}D\right) = \log_2 {\left\|D\psi^B\right\|}_\infty\\
        &\geqq\min_{Z^{AB}} \log_2{\left\|Z^B\right\|}_\infty ={{H_{\max}(A|B)}_\psi}.
      \end{split}
    \]
\end{IEEEproof}

It is natural to ask how tight our converse bounds in Theorem~\ref{thm:new} and Corollary~\ref{col:tractable_converse} are.
In the following analysis of the tightness, we consider non-catalytic exact state merging using one-way LOCC~\cite{C7} from $A$ to $B$ for simplicity,
and we use the following proposition.
\begin{proposition}
\label{prp:equivalence}
    \textit{A necessary and sufficient condition for non-catalytic exact state merging by one-way LOCC\@.}
    Given any pure state $\Ket{\psi}^{RAB}$ satisfying $\psi^R=\frac{\mathbbm{1}^R}{D}$,
    there exists one-way LOCC map $\mathcal{M}^{A\to B}$ from $A$ to $B$ achieving
    \[
        \id^R\otimes\mathcal{M}^{A\to B}\left(\psi^{RAB}\otimes{\Phi_K^+}^{\overline{A}\overline{B}}\right)=\psi^{RB'B}
    \]
    if and only if
    there exists a mixed-unitary channel $\mathcal{U}(\rho)=\sum_m p\left(m\right) U_m\rho U_m^\dag$~\cite{W11}, where $p\left(m\right)$ is a probability distribution and $U_m$ for each $m$ is a unitary, achieving
    \begin{equation}
      \label{eq:mixed_unitary}
      \id^R\otimes\mathcal{U}^{\hat{B}}\left({\Phi_D^+}^{R\hat{B}}\right)=\psi^{RB}\otimes\frac{\mathbbm{1}_K^{\overline{B}}}{K},
    \end{equation}
    where $\mathcal{H}^{\hat{B}}=\mathcal{H}^{B}\otimes\mathcal{H}^{\overline{B}}$, and $\Ket{\Phi_D^+}^{R\hat{B}}\coloneqq\frac{1}{\sqrt{D}}\sum_{l=0}^{D-1}\Ket{l}^R\otimes\Ket{l}^{\hat{B}}$.
\end{proposition}

\begin{IEEEproof}
  \textit{If part:} Assume that
  \[
    \psi^{RB}\otimes\frac{\mathbbm{1}_K^{\overline{B}}}{K}=\sum_m p\left(m\right)\left(\mathbbm{1}^R\otimes U_m^{\hat{B}}\right){\Phi_D^+}^{R{\hat{B}}}{\left(\mathbbm{1}^R\otimes U_m^{\hat{B}}\right)}^\dag.
  \]
  A purification yields
  \[
    \begin{split}
      &\left(\mathbbm{1}^{RB \overline{B}}\otimes U \right)\left(\Ket{\psi}^{RAB}\otimes\Ket{\Phi_K^+}^{\overline{A}\overline{B}}\right)\\
      &=\sum_m \sqrt{p\left(m\right)}\Ket{m}^{A_0}\otimes\left(\mathbbm{1}^R\otimes U_m^{\hat{B}}\right)\Ket{\Phi_D^+}^{R{\hat{B}}},
    \end{split}
  \]
  where $\mathcal{H}^{A_0}$ is $A$'s auxiliary system, and $U$ is an isometry performed by $A$.
  Hence, a one-way LOCC map from $A$ to $B$ represented by ${\left\{\left(\Bra{m}^{A_0} U\right)\otimes {\left(U_m^{\hat{B}}\right)}^\dag\right\}}_m$, where the post-measurement state of $A$ is traced out, achieves, for each $m$,
  \[
    \begin{split}
      &\mathbbm{1}^{R}\otimes\left[\left(\Bra{m}^{A_0} U\right)\otimes {\left(U_m^{\hat{B}}\right)}^\dag\right]\left(\Ket{\psi}^{RAB}\otimes\Ket{\Phi_K^+}^{\overline{A}\overline{B}}\right)\\
      &\propto\Ket{\Phi_D^+}^{R{\hat{B}}},
    \end{split}
  \]
  and $\Ket{\Phi_D^+}^{R{\hat{B}}}$ on the right-hand side can be transformed into $\Ket{\psi}^{RB^\prime B}$ by $B$'s local isometry.

  \textit{Only if part:} Assume that there exists $A$'s positive operator-valued measure (POVM~\cite{W5}) ${\left\{\Lambda_m\right\}}_m$ on $\mathcal{H}^A\otimes\mathcal{H}^{\overline{A}}$ satisfying for each $m$
  \[
    \begin{split}
      &\tr_A\left[\left(\mathbbm{1}^{RB\overline{B}}\otimes\Lambda_m\right)\left(\psi^{RAB}\otimes{\Phi_K^+}^{\overline{A}\overline{B}}\right)\right]\\
      &=p\left(m\right){\left(\mathbbm{1}^{R}\otimes U_m^{\hat{B}}\right)}{\Phi_D^+}^{R{\hat{B}}}{\left(\mathbbm{1}^{R}\otimes U_m^{\hat{B}}\right)}^\dag,
    \end{split}
  \]
  where $p\left(m\right)$ is a probability distribution, and $U_m^{\hat{B}}$ is $B$'s unitary correction conditioned by $m$.
  Note that ${\Phi_D^+}^{R{\hat{B}}}$ on the right-hand side can be transformed into $\Ket{\psi}^{RB^\prime B}$ by $B$'s local isometry.
  Then, we obtain
  \[
    \begin{split}
      &\psi^{RB}\otimes\frac{\mathbbm{1}_K^{\overline{B}}}{K}\\
      &=\sum_m\tr_A\left[\left(\mathbbm{1}^{RB\overline{B}}\otimes\Lambda_m\right)\left(\psi^{RAB}\otimes{\Phi_K^+}^{\overline{A}\overline{B}}\right)\right]\\
      &=\sum_m p\left(m\right)\left(\mathbbm{1}^R\otimes U_m^{\hat{B}}\right){\Phi_D^+}^{R{\hat{B}}}{\left(\mathbbm{1}^R\otimes U_m^{\hat{B}}\right)}^\dag\\
      &=\id^R\otimes\mathcal{U}^{\hat{B}}\left({\Phi_D^+}^{R\hat{B}}\right).
    \end{split}
  \]
\end{IEEEproof}

Note that it is straightforward to generalize the above proof of Proposition~\ref{prp:equivalence} on non-catalytic exact state merging to the catalytic setting, that is,
\begin{align*}
  &\id^R\otimes\mathcal{M}^{A\to B}\left(\psi^{RAB}\otimes{\Phi_K^+}^{\overline{A}\overline{B}}\right)={\psi}^{RB^\prime B}\otimes{\Phi_L^+}^{\overline{A}\overline{B}}\\
  &\Leftrightarrow\id^R\otimes\mathcal{U}^{\hat{B}}\left({\Phi_D^+}^{R{\hat{B}}}\otimes\frac{\mathbbm{1}_L^{\hat{B}}}{L}\right)=\psi^{RB}\otimes\frac{\mathbbm{1}_K^{\overline{B}}}{K},
\end{align*}
which can also be shown for quantum state redistribution in the approximate scenarios~\cite{B13,B15}.

For qubits, our converse bound in Corollary~\ref{col:tractable_converse} is tight enough to provide the optimal entanglement cost as shown in the following.
Note that an equivalent condition in terms of Schmidt coefficients of $\Ket{\psi_l}^{AB}$ in Eq.~\eqref{eq:max} is also shown in Theorem~II.1.\ in Ref.~\cite{O}.

\begin{theorem}
\label{thm:qubit}
    \textit{Optimal entanglement cost of non-catalytic exact state merging for qubits.}
    Consider any three-qubit pure state $\Ket{\psi}^{RAB}\in{\left(\mathbb{C}^2\right)}^{\otimes 3}$ satisfying $\psi^R=\frac{\mathbbm{1}^R}{2}$,
    non-catalytic exact state merging of $\Ket{\psi}^{RAB}$ is achievable if and only if
    \[
        \log_2 K \geqq \log_2 \left\lceil\lambda_0^B D\right\rceil,
    \]
    where the notations are the same as those in Corollary~\ref{col:tractable_converse}.
    Equivalently,
    non-catalytic exact state merging of $\Ket{\psi}^{RAB}$ where $\psi^R=\frac{\mathbbm{1}^R}{2}$ is achievable at entanglement cost $\log_2 K = 0$ if and only if $\psi^B=\frac{\mathbbm{1}^B}{2}$,
    and otherwise entanglement cost $\log_2 K = 1$ is required.
\end{theorem}

\begin{IEEEproof}
    \textit{If part:}
    We assume that $\psi^B=\frac{\mathbbm{1}^B}{2}$ and show the existence of an LOCC algorithm for exact state merging of $\Ket{\psi}^{RAB}$ achieving $\log_2 K=0$
    since otherwise quantum teleportation of $\psi^A$ achieves $\log_2 K = 1$.
    To show the existence of the LOCC algorithm, Proposition~\ref{prp:equivalence} implies that it is sufficient to prove the existence of a mixed-unitary channel $\mathcal{U}$ achieving
    \begin{equation}
      \label{eq:qubit}
      \id^R\otimes\mathcal{U}^B\left({\Phi_2^+}^{RB}\right)=\psi^{RB}.
    \end{equation}
    Note that $\mathcal{H}^{\hat{B}}$ in Eq.~\eqref{eq:mixed_unitary} in Proposition~\ref{prp:equivalence} is simply written as $\mathcal{H}^B$ in Eq.~\eqref{eq:qubit}, since $\mathcal{H}^{\hat{B}}=\mathcal{H}^B$ in this proof.

    Given $\psi^{RB}$ where $\psi^R=\frac{\mathbbm{1}^R}{2}$,
    we can regard $\psi^{RB}$ as a normalized operator of the Choi operator~\cite{W11} of a CPTP map $\mathcal{U}^B$.
    Tracing out $\mathcal{H}^R$ for $\psi^{RB}$ yields
    \[
        \mathcal{U}^B\left(\frac{\mathbbm{1}^B}{2}\right)=\psi^B=\frac{\mathbbm{1}^B}{2},
    \]
    that is, $\mathcal{U}^B$ is unital.
    Since any unital channel on a qubit is a mixed-unitary channel~\cite{W11}, $\mathcal{U}^B$ is a mixed-unitary channel, which yield the conclusion.
\end{IEEEproof}

As for qudits of more than two dimension, our converse bound in Theorem~\ref{thm:new} is not necessarily achievable, since there exists an example of non-catalytic exact state merging which does not satisfy the equality of~\eqref{eq:lower_non_catalytic}.
We show a three-qutrit state of which any one-way LOCC algorithm for non-catalytic exact state merging fails to achieve
\[
    \log_2 K =\min\left\{\log_2 K: \frac{\mathbbm{1}_K}{K}\otimes\psi^{B}\prec\psi^{AB}\right\}.
\]

\begin{proposition}
\label{prp:qutrit}
  \textit{Impossibility of achieving the converse bound of entanglement cost of non-catalytic exact state merging for qutrits.}
  There exists a three-qutrit pure state $\Ket{\psi}^{RAB}\in{\left(\mathbb{C}^3\right)}^{\otimes 3}$ satisfying $\psi^R=\frac{\mathbbm{1}^R}{D}$ where $D=3$, such that non-catalytic exact state merging of $\Ket{\psi}^{RAB}$ cannot be achieved by any one-way LOCC algorithm at entanglement cost
  \[
    \log_2 K =\min\left\{\log_2 K: \frac{\mathbbm{1}_K}{K}\otimes\psi^{B}\prec\psi^{AB}\right\},
  \]
  where the notations are the same as those in Theorem~\ref{thm:new}.
\end{proposition}

\begin{IEEEproof}
    Consider a CPTP map
    \[
      \mathcal{N}(\rho)=\frac{1}{2}(\tr\rho)\mathbbm{1}-\frac{1}{2}\rho^\textup{T},
    \]
    where $\rho^\textup{T}$ is transpose of $\rho$ with respect to the computational basis.
    The Choi operator of $\mathcal{N}$ is written as
    \[
        \begin{split}
          &J(\mathcal{N})\coloneqq\\
          &\frac{1}{2}\left(\Ket{2}\otimes\Ket{1}-\Ket{1}\otimes\Ket{2}\right){\left(\Bra{2}\otimes\Bra{1}-\Bra{1}\otimes\Bra{2}\right)}+\\
          &\frac{1}{2}\left(\Ket{0}\otimes\Ket{2}-\Ket{2}\otimes\Ket{0}\right){\left(\Bra{0}\otimes\Bra{2}-\Bra{2}\otimes\Bra{0}\right)}+\\
          &\frac{1}{2}\left(\Ket{1}\otimes\Ket{0}-\Ket{0}\otimes\Ket{1}\right){\left(\Bra{1}\otimes\Bra{0}-\Bra{0}\otimes\Bra{1}\right)}.
        \end{split}
    \]
    This map $\mathcal{N}$ is a unital channel but not a mixed-unitary channel~\cite{L4,W11}.

    Consider
    \[
        \psi^{RB}=\frac{J(\mathcal{N})}{3}.
    \]
    A purification of $\psi^{RB}$ is
    \[
        \begin{split}
          &\Ket{\psi}^{RAB}=\\
          &\frac{1}{\sqrt{3}}\Ket{0}^A\otimes{\left(\frac{1}{\sqrt{2}}\Ket{2}^R\otimes\Ket{1}^B-\frac{1}{\sqrt{2}}\Ket{1}^R\otimes\Ket{2}^B\right)}+\\
          &\frac{1}{\sqrt{3}}\Ket{1}^A\otimes{\left(\frac{1}{\sqrt{2}}\Ket{0}^R\otimes\Ket{2}^B-\frac{1}{\sqrt{2}}\Ket{2}^R\otimes\Ket{0}^B\right)}+\\
          &\frac{1}{\sqrt{3}}\Ket{2}^A\otimes{\left(\frac{1}{\sqrt{2}}\Ket{1}^R\otimes\Ket{0}^B-\frac{1}{\sqrt{2}}\Ket{0}^R\otimes\Ket{1}^B\right)}
        \end{split}
    \]
    For this state, it holds that
    \[
        \begin{split}
            &\psi^R=\frac{\mathbbm{1}^R}{3},\\
            &\psi^B=\frac{\mathbbm{1}^B}{3}.
        \end{split}
    \]
    Hence, we obtain
    \[
        \min\left\{\log_2 K: \frac{\mathbbm{1}_K}{K}\otimes\psi^{B}\prec\psi^{AB}\right\}=0.
    \]

    We assume that there exists a one-way LOCC algorithm for non-catalytic exact state merging of $\Ket{\psi}^{RAB}$ at entanglement cost $\log_2 K=0$ to derive a contradiction.
    Due to Proposition~\ref{prp:equivalence}, this assumption is equivalent to the existence of a mixed-unitary channel $\mathcal{U}$ such that
    \[
        \id^R\otimes\mathcal{U}^B\left({\Phi_3^+}^{RB}\right)=\psi^{RB}=\frac{J(\mathcal{N})}{3},
    \]
    where, in the same way as Eq.~\eqref{eq:qubit}, $\mathcal{H}^{\hat{B}}$ in Eq.~\eqref{eq:mixed_unitary} in Proposition~\ref{prp:equivalence} is written as $\mathcal{H}^B$.
    Therefore, $\mathcal{N}=\mathcal{U}$ is necessary, which contradicts to the fact that $\mathcal{N}$ is not a mixed-unitary channel, and we obtain the conclusion.
\end{IEEEproof}

\section{\label{sec:examples}Implications}
We discuss implications of our main results. In the following, we omit $\otimes$ in representing states.
We define
\[
    \begin{split}
        \Ket{+}&\coloneqq\frac{1}{\sqrt{2}}\left(\Ket{0}+\Ket{1}\right),\\
        \Ket{\Psi^\pm}&\coloneqq\frac{1}{\sqrt{2}}\left(\Ket{0}\Ket{1}\pm\Ket{1}\Ket{0}\right),\\
        \Ket{\Phi^\pm}&\coloneqq\frac{1}{\sqrt{2}}\left(\Ket{0}\Ket{0}\pm\Ket{1}\Ket{1}\right).
    \end{split}
\]

\begin{implication}
\label{ex:1}
\textit{Reduced entanglement cost in exact state merging compared with quantum teleportation and exact state splitting, by performing a measurement on the classical part followed by classical communication.}
Consider a tripartite Greenberger-Horne-Zeilinger (GHZ) state of $d$-dimensional systems for any $d\geqq 2$
\[
  \Ket{\textup{GHZ}_d}^{RAB}\coloneqq\frac{1}{\sqrt{d}}\sum_{l=0}^{d-1}\Ket{l}^R\Ket{l}^A\Ket{l}^B.
\]
Quantum teleportation of $A$'s part of $\Ket{\textup{GHZ}_d}^{RAB}$ on $A$ requires $\log_2 d$ ebits, that is, $\Ket{\Phi_d^+}$ for an initial resource state.
Note that exact state splitting analyzed in Appendix~\ref{sec:split} also requires $\log_2 d$ ebits, as shown in Theorem~\ref{thm:split} in Appendix~\ref{sec:split_result}.
By contrast, the algorithms for exact state merging of $\Ket{\textup{GHZ}_d}^{RAB}$ in Theorems~\ref{thm:merge} and~\ref{thm:merge_without_catalyst} achieve $\log_2 K - \log_2 L = 0<\log_2 d$ and $\log_2 K  = 0<\log_2 d$, respectively.
In a similar way, our accompanying paper~\cite{Y13} shows that our algorithms can be used for achieving \textit{zero} entanglement cost in exact state merging of multipartite code states of quantum error correcting codes, where these code states are regarded as tripartite states.
\end{implication}

\begin{implication}
\label{ex:3}
\textit{Negative entanglement cost in exact state merging by entanglement distillation from the redundant part.}
Consider a pure state
\[
  \begin{split}
    \Ket{\psi}^{RAB}=\frac{1}{\sqrt{3}}\Big(&\Ket{0}^R\Ket{\Psi^+}^{A_1B_1}\Ket{\Phi^-}^{A_2B_2}\Ket{\Phi^+}^{A_3B_3}+\\
                                            &\Ket{1}^R\Ket{0}^{A_1}\Ket{0}^{B_1}\Ket{\Phi^-}^{A_2B_2}\Ket{\Phi^+}^{A_3B_3}+\\
                                            &\Ket{2}^R\Ket{2}^{A_1}\Ket{2}^{B_1}\Ket{0}^{A_2}\Ket{0}^{B_2}\Ket{\Psi^-}^{A_3B_3}\Big),
  \end{split}
\]
where each of $\mathcal{H}^A=\mathcal{H}^{A_1}\otimes\mathcal{H}^{A_2}\otimes\mathcal{H}^{A_3}$ and $\mathcal{H}^B=\mathcal{H}^{B_1}\otimes\mathcal{H}^{B_2}\otimes\mathcal{H}^{B_3}$ is of $3\times 2\times 2=12$ dimension.
Quantum teleportation of $\psi^A$ requires $\log_2 12$ ebits, that is, $\Ket{\Phi_{12}^+}$ for an initial resource state.
By contrast, the algorithms for exact state merging of $\Ket{\psi}^{RAB}$ in Theorems~\ref{thm:merge} and~\ref{thm:merge_without_catalyst} achieve $\log_2 K - \log_2 L = -1 < 0$ and $\log_2 K = 0$, respectively.
The former negative entanglement cost leads to a net gain of shared entanglement.
\end{implication}

\begin{implication}
\label{ex:2}
\textit{Improvement in converse bounds of entanglement cost in exact state merging.}
Consider a three-qubit pure state
\[
    \Ket{\psi}^{RAB}=\frac{1}{\sqrt{2}}\Big(\Ket{0}^R\Ket{\Psi^+}^{AB} +\Ket{1}^R\Ket{0}^A\Ket{0}^{B}\Big).
\]
The algorithms for exact state merging of $\Ket{\psi}^{RAB}$ in Theorems~\ref{thm:merge} and~\ref{thm:merge_without_catalyst} require $\log_2 K - \log_2 L = 1$ and $\log_2 K = 1$, respectively.
Since $\psi^B\neq\frac{\mathbbm{1}^B}{2}$,
the latter equality for non-catalytic exact state merging is optimal due to Theorem~\ref{thm:qubit}.
As for the former in the catalytic setting, this example shows the difference between the converse bounds of entanglement cost of exact state merging in Corollary~\ref{col:tractable_converse} and Lemma~\ref{lem:old}.
In this case,
\begin{align*}
  &\log_2 \left({\lambda_0^B}D\right)=\log_2 \frac{3}{2} > 0.5849,\\
  &{H_{\max}(A|B)}_\psi < 0.5432,
\end{align*}
where the notations are the same as those in Theorem~\ref{thm:new} and Lemma~\ref{lem:old},
and the value of ${H_{\max}(A|B)}_\psi$ is calculated by a semidefinite programming~\cite{V2} using Split Conic Solver (SCS)~\cite{S8} and YALMIP~\cite{L5}.
These calculations imply that our converse bounds in Theorem~\ref{thm:new} and Corollary~\ref{col:tractable_converse} can be strictly tighter than the existing converse bound obtained from Lemma~\ref{lem:old}.
\end{implication}

\begin{implication}
\label{ex:4}
\textit{Asymmetry between $A$ and $B$ in exact state merging.}
Consider a three-qubit pure state
\[
    \Ket{\psi}^{RAB}=\frac{1}{\sqrt{2}}\Big(\Ket{0}^R\Ket{0}^{A}\Ket{0}^{B} +\Ket{1}^R\Ket{1}^A\Ket{+}^{B}\Big).
\]
The algorithms for exact state merging of $\Ket{\psi}^{RAB}$ in Theorems~\ref{thm:merge} and~\ref{thm:merge_without_catalyst} require $\log_2 K - \log_2 L = 1$ and $\log_2 K = 1$, respectively.
Since $\psi^B\neq\frac{\mathbbm{1}^B}{2}$, the latter equality for non-catalytic exact state merging is optimal due to Theorem~\ref{thm:qubit}.

In contrast, interchange $A$ and $B$ for $\Ket{\psi}^{RAB}$ to consider
\[
    \Ket{\psi^\prime}^{RAB}=\frac{1}{\sqrt{2}}\Big(\Ket{0}^R\Ket{0}^{A}\Ket{0}^{B} +\Ket{1}^R\Ket{+}^A\Ket{1}^{B}\Big).
\]
In the same way as the above case of $\Ket{\psi}^{RAB}$,
the algorithms for exact state merging of $\Ket{\psi^\prime}^{RAB}$ in Theorems~\ref{thm:merge} and~\ref{thm:merge_without_catalyst} require $\log_2 K - \log_2 L = 1$ and $\log_2 K = 1$, respectively.
However, since $\psi^B=\frac{\mathbbm{1}^B}{2}$, Theorem~\ref{thm:qubit} implies that there exists an algorithm
for non-catalytic exact state merging of $\Ket{\psi^\prime}^{RAB}$ achieving $\log_2 K = 0 < 1$.
Indeed, $\Ket{\psi^\prime}^{RAB}$ can also be written as
\[
    \begin{split}
      \Ket{\psi^\prime}^{RAB}=&\sqrt{\frac{1}{2}+\frac{\sqrt{2}}{4}}{\left[\frac{\left(1+\sqrt{2}\right)\Ket{0}+\Ket{1}}{\sqrt{4+2\sqrt{2}}}\right]}^A\Ket{\Phi^-}^{RB}+\\
                       &\sqrt{\frac{1}{2}-\frac{\sqrt{2}}{4}}{\left[\frac{\left(1-\sqrt{2}\right)\Ket{0}+\Ket{1}}{{\sqrt{4-2\sqrt{2}}}}\right]}^A\Ket{\Phi^+}^{RB},
    \end{split}
\]
and hence, $A$'s measurement in basis
\[
    \left\{\frac{\left(1+\sqrt{2}\right)\Ket{0}+\Ket{1}}{\sqrt{4+2\sqrt{2}}}, \frac{\left(1-\sqrt{2}\right)\Ket{0}+\Ket{1}}{{\sqrt{4-2\sqrt{2}}}}\right\}
\]
yields a maximally entangled state between $R$ and $B$.

These cases imply that the difference in entanglement costs between the optimal algorithm and the algorithms presented in Theorems~\ref{thm:merge} and~\ref{thm:merge_without_catalyst} may arise depending on whether the quantum part of the Koashi-Imoto decomposition can be merged at less entanglement cost than performing quantum teleportation.
Note that the optimal algorithm obtained in Theorem~\ref{thm:qubit} works only for qubits, and Proposition~\ref{prp:qutrit} implies that extension to qudits is not straightforward.
\end{implication}

\begin{implication}
\label{ex:5}
\textit{Special cases where the achievability and converse bounds for exact state merging coincide.}
We discuss special cases where one of the subsystems of system $\mathcal{H}^R\otimes\mathcal{H}^A\otimes\mathcal{H}^B$ for a given state $\Ket{\psi}^{RAB}$ is initially decoupled from the others.
In these cases, the achievability bound for exact state merging in Theorem~\ref{thm:merge} coincides with the converse bound in Theorem~\ref{thm:new}.
Note that in general, there may exist a gap between these bounds as discussed in Implications~\ref{ex:2} and~\ref{ex:4}, while full characterization of the cases where this gap closes is unknown.

Consider the case where system $\mathcal{H}^R$ is initially decoupled with the others, and a given pure state is in the form of
\[
  \Ket{\psi_{R\textup{-}AB}}^{RAB}=\Ket{\mu}^{R}\otimes\Ket{\nu}^{AB}.
\]
Due to the Koashi-Imoto decomposition of $\Ket{\psi}^{RAB}$ in Lemma~\ref{lem:koashi_imoto_decomposition_tripartite}, we obtain the decomposition of $\mathcal{H}^A$
\[
  \begin{split}
    \mathcal{H}^A&=\mathcal{H}^{a_0^L},
  \end{split}
\]
where in terms of the notations of Lemma~\ref{lem:koashi_imoto_decomposition_tripartite}, $J=1$, and $\mathcal{H}^{a_0^R}$ does not explicitly appear since in this case
\[
  \begin{split}
    \dim\mathcal{H}^{a_0^L}&=\dim\mathcal{H}^A,\quad\dim\mathcal{H}^{a_0^R}=1.
  \end{split}
\]
As for $\Ket{\psi_{R\textup{-}AB}}^{RAB}$, the decomposition yields
\[
  \Ket{\psi_{R\textup{-}AB}}^{RAB}=\Ket{\mu}^{R}\otimes\Ket{\nu}^{a_0^L b_0^L},
\]
and we define
\[
  \lambda_0\coloneqq\lambda_0^{a_0^L}=\lambda_0^B,
\]
where the notations are the same as those in Theorems~\ref{thm:merge} and~\ref{thm:new}.
The algorithm in Theorem~\ref{thm:merge} for exact state merging of $\Ket{\psi_{R\textup{-}AB}}^{RAB}$ achieves for any $\delta > 0$
\[
  \log_2 K - \log_2 L \leqq \log_2\lambda_0 + \delta,
\]
where shared entanglement is distilled by Subprocess~1 in the proof of Theorems~\ref{thm:merge}.
The converse bound in Theorem~\ref{thm:new} shows for any algorithm for exact state merging of $\Ket{\psi_{R\textup{-}AB}}^{RAB}$
\[
  \log_2 K - \log_2 L \geqq \log_2\lambda_0.
\]

Next, consider the case where system $\mathcal{H}^B$ is initially decoupled with the others, and a given pure state is in the form of
\[
  \Ket{\psi_{B\textup{-}RA}}^{RAB}=\Ket{\mu}^{B}\otimes\Ket{\nu}^{RA}.
\]
Due to the Koashi-Imoto decomposition of $\Ket{\psi_{B\textup{-}RA}}^{RAB}$ in Lemma~\ref{lem:koashi_imoto_decomposition_tripartite}, we obtain the decomposition of $\mathcal{H}^A$
\[
  \begin{split}
    \mathcal{H}^A&=\mathcal{H}^{a_0^R}\oplus\mathcal{H}^{a_1^L},
  \end{split}
\]
where in terms of the notations of Lemma~\ref{lem:koashi_imoto_decomposition_tripartite}, $J=1$, and $\mathcal{H}^{a_0^L}$ and $\mathcal{H}^{a_1^R}$ do not explicitly appear since in this case
\begin{align*}
  \dim\mathcal{H}^{a_0^L}&=1,\\
  \dim\mathcal{H}^{a_0^R}&=\rank\psi_{B\textup{-}RA}^A,\\
  \dim\mathcal{H}^{a_1^L}&=\dim\mathcal{H}^A-\rank\psi_{B\textup{-}RA}^A,\\
  \dim\mathcal{H}^{a_1^R}&=1.
\end{align*}
As for $\Ket{\psi_{B\textup{-}RA}}^{RAB}$, the decomposition yields
\[
  \Ket{\psi_{B\textup{-}RA}}^{RAB}=\Ket{\mu}^{b_0^L}\otimes\Ket{\nu}^{R a_0^R}.
\]
The algorithm in Theorem~\ref{thm:merge} for exact state merging of $\Ket{\psi_{B\textup{-}RA}}^{RAB}$ achieves
\[
  \log_2 K = \rank \nu^{a_0^R} = \rank\psi_{B\textup{-}RA}^A,\quad \log_2 L = 0.
\]
where $\nu^{a_0^R}$ is transferred using quantum teleportation in Subprocess~2 in the proof of Theorems~\ref{thm:merge}.
The converse bound in Theorem~\ref{thm:new} shows for any algorithm for exact state merging of $\Ket{\psi_{B\textup{-}RA}}^{RAB}$
\[
  \log_2 K - \log_2 L \geqq \rank\psi_{B\textup{-}RA}^A.
\]

Finally, consider the case where system $\mathcal{H}^A$ is initially decoupled with the others, and a given pure state is in the form of
\[
  \Ket{\psi_{A\textup{-}RB}}^{RAB}=\Ket{\mu}^{A}\otimes\Ket{\nu}^{RB}.
\]
Due to the Koashi-Imoto decomposition of $\Ket{\psi_{A\textup{-}RB}}^{RAB}$ in Lemma~\ref{lem:koashi_imoto_decomposition_tripartite}, we obtain the decomposition of $\mathcal{H}^A$
\[
  \begin{split}
    \mathcal{H}^A&=\mathcal{H}^{a_0^L},
  \end{split}
\]
where in terms of the notations of Lemma~\ref{lem:koashi_imoto_decomposition_tripartite}, $J=1$, and $\mathcal{H}^{a_0^R}$ does not explicitly appear since in this case
\begin{align*}
  \dim\mathcal{H}^{a_0^L}=\dim\mathcal{H}^A,\quad \dim\mathcal{H}^{a_0^R}=1.
\end{align*}
As for $\Ket{\psi_{A\textup{-}RB}}^{RAB}$, the decomposition yields
\[
  \Ket{\psi_{A\textup{-}RB}}^{RAB}=\Ket{\mu}^{a_0^L}\otimes\Ket{\nu}^{R b_0^R}.
\]
The algorithm in Theorem~\ref{thm:merge} for exact state merging of $\Ket{\psi_{A\textup{-}RB}}^{RAB}$ achieves
\[
  \log_2 K = \log_2 L = 0,
\]
where $B$ locally prepares a state corresponding to $\Ket{\mu}^{a_0^L}$ due to Subprocess~3 in the proof of Theorems~\ref{thm:merge}.
The converse bound in Theorem~\ref{thm:new} shows for any algorithm for exact state merging of $\Ket{\psi_{A\textup{-}RB}}^{RAB}$
\[
  \log_2 K - \log_2 L \geqq 0.
\]
\end{implication}

\section{\label{sec:conclusion}Conclusion}
We constructed exact algorithms for one-shot state merging, which work for any state of an arbitrarily small-dimensional system and satisfy arbitrarily high fidelity requirements.
The algorithms retain the essential feature of state merging; that is, entanglement cost can be reduced by exploiting a structure of a given state.
This feature arises because the Koashi-Imoto decomposition of the given state shows the classical part, the nonclassical (quantum) part, and the redundant part of the state, and the redundant part can be used for entanglement distillation,
while the classical part can be merged by a measurement followed by classical communication of the measurement outcome.
In these algorithms, it is crucial to coherently combine different subprocesses, namely, entanglement distillation from the redundant part and quantum teleportation of the quantum part, using controlled measurements and controlled isometries.
In addition to achievability bounds for an arbitrarily small-dimensional system derived from the algorithms, we provided an improved converse bound of entanglement cost in exact state merging, which is proven to be optimal when a purification of the given state to be merged is a three-qubit state,
while further research will be needed to establish a general optimal strategy for achieving exact state merging.
As shown in Appendix~\ref{sec:approximate}, these results on exact state merging can also be extended to their approximate versions by means of smoothing~\cite{R2,T5}, while exact state merging suffices to reduce entanglement cost in cases relevant to distributed quantum information processing, such as code states of quantum error correcting codes, as discussed in Remark~\ref{remark:approximate} in Sec.~\ref{sec:merge}.

Our results complement existing algorithms for one-shot state merging achieving near optimality on a large scale, opening the way to another direction for future research on small and intermediate scales.
As investigated in our accompanying paper~\cite{Y13},
the algorithms in this paper serve as essential tools for analyzing exact transformation of a multipartite entangled state shared among spatially separated parties connected by a communication network.
We leave further investigation of application of our results for future work.

\appendices%

\section{\label{sec:approximate}Approximate state merging for arbitrarily small-dimensional systems}

In this appendix, we extend our results on exact state merging presented in Sec.~\ref{sec:result} in the main text to their approximate versions, by means of smoothing~\cite{R2,T5}.
We consider the catalytic setting, while extension of our results on non-catalytic exact state merging is also possible in the same way.
Note that while allowing small error in smoothing may provide better bounds, the bounds obtained by smoothing usually include optimization over a ball of close states, and exact state merging already suffices for useful examples including those relevant to distributed quantum information processing, as discussed in Remark~\ref{remark:approximate} in Sec.~\ref{sec:merge} in the main text.
In the following, after defining the task of approximate state merging in Sec.~\ref{sec:def_approximate}, we provide an achievability bound of entanglement cost in approximate state merging in Sec.~\ref{sec:achievability_approximate} and also analyze a converse bound in Sec.~\ref{sec:converse_approximate}.

\subsection{\label{sec:def_approximate}Definition of approximate state merging}

The task of approximate state merging is defined as follows.
\begin{definition}
  \textit{Approximate state merging.}
  Approximate state merging of a purified given state $\Ket{\psi}^{RAB}$ within a given error $\epsilon\geqq 0$ is a task for parties $A$ and $B$ to achieve
  \[
    \begin{split}
      &F^2\left(\id^R\otimes\tilde{\mathcal{M}}\left({\psi}^{RAB}\otimes{\Phi^+_K}^{\overline{A}\overline{B}}\right),{\psi}^{RB'B}\otimes{\Phi^+_L}^{\overline{A}\overline{B}}\right)\\
      &\geqq 1-\epsilon^2,
    \end{split}
  \]
  where $\tilde{\mathcal{M}}:\mathcal{B}\left(\mathcal{H}^A\otimes\mathcal{H}^B\otimes\mathcal{H}^{\overline{A}}\otimes\mathcal{H}^{\overline{B}}\right)\to\mathcal{B}\left(\mathcal{H}^{B'}\otimes\mathcal{H}^B\otimes\mathcal{H}^{\overline{A}}\otimes\mathcal{H}^{\overline{B}}\right)$ is an LOCC map, which can be constructed depending on the classical description of $\Ket{\psi}^{RAB}$, and ${F^2\left(\rho,\sigma\right)}\coloneqq{\left({\left\|\sqrt{\rho}\sqrt{\sigma}\right\|}_1\right)}^2$ is the fidelity.
  The entanglement cost of an algorithm for approximate state merging is defined as $\log_2 K-\log_2 L$.
\end{definition}

\subsection{\label{sec:achievability_approximate}Achievability bound for approximate state merging applicable to arbitrarily small-dimensional systems}
Given any pure state $\Ket{\psi}^{RAB}$ and an error $\epsilon\geqq 0$,
we extend Theorem~\ref{thm:merge} and obtain an achievability bound of entanglement cost of approximate state merging of $\Ket{\psi}^{RAB}$ within $\epsilon$ as follows.
Consider the Koashi-Imoto decomposition of any normalized pure state $\Ket{\tilde\psi}^{RAB}$ satisfying ${F^2\left(\Ket{\psi}\Bra{\psi},\Ket{\tilde\psi}\Bra{\tilde\psi}\right)}\geqq 1-{\left(\frac{\epsilon}{2}\right)}^2$.
Due to Lemma~\ref{lem:koashi_imoto_decomposition_tripartite} on the Koashi-Imoto decomposition, $\mathcal{H}^A$ and $\supp\left({\tilde\psi}^{B}\right)$ are uniquely decomposed into
\begin{align}
\label{eq:notation_space_approx}
  \mathcal{H}^A=\bigoplus_{j=0}^{J-1}{\mathcal{H}}^{\tilde{a}_j^L}\otimes{\mathcal{H}}^{\tilde{a}_j^R},\quad
  \supp\left({\tilde\psi}^{B}\right)=\bigoplus_{j=0}^{J-1}{\mathcal{H}}^{\tilde{b}_j^L}\otimes{\mathcal{H}}^{\tilde{b}_j^R},
\end{align}
and $\Ket{\tilde\psi}^{RAB}$ is uniquely decomposed into
\begin{equation}
\label{eq:notation_state_approx}
  \Ket{\tilde\psi}^{RAB}=\bigoplus_{j=0}^{J-1}\sqrt{\tilde{p}\left(j\right)}\Ket{\omega_j}^{\tilde{a}_j^L \tilde{b}_j^L}\otimes\Ket{\phi_j}^{R \tilde{a}_j^R \tilde{b}_j^R},
\end{equation}
where $\tilde{p}\left(j\right)$ is a probability distribution.
Using these notations, Theorem~\ref{thm:merge} on exact state merging can be extended to approximate state merging as follows.
\begin{theorem}
\label{thm:approximate}
  \textit{An achievability bound of entanglement cost of approximate state merging applicable to arbitrarily small-dimensional systems.}
  Given any pure state $\Ket{\psi}^{RAB}$,
  any $\epsilon \geqq 0$,
  and any $\delta > 0$,
  there exists an algorithm for approximate state merging of $\Ket{\psi}^{RAB}$ within $\epsilon$ achieving
  \begin{equation}
    \label{eq:approximate}
    \begin{split}
      &\log_2 K-\log_2 L\\
      &\leqq \min_{\Ket{\tilde\psi}}\max_{j}\left\{\log_2\left(\lambda_0^{\tilde{a}_j^L}\dim\mathcal{H}^{\tilde{a}_j^R}\right)\right\} + \delta,
    \end{split}
  \end{equation}
  where the notations are the same as those in Eqs.~\eqref{eq:notation_space_approx} and~\eqref{eq:notation_state_approx}, $\lambda^{\tilde{a}_j^L}_0$ is the largest eigenvalue of $\omega_j^{\tilde{a}_j^L}$, and the minimization is over any normalized pure state $\Ket{\tilde\psi}^{RAB}$ satisfying ${F^2\left(\Ket{\psi}\Bra{\psi},\Ket{\tilde\psi}\Bra{\tilde\psi}\right)}\geqq 1-{\left(\frac{\epsilon}{2}\right)}^2$.
\end{theorem}

\begin{IEEEproof}
  We show that the LOCC map $\tilde{\mathcal{M}}$ for exact state merging of the approximate state $\Ket{\tilde\psi}$ providing the minimum in Eq.~\eqref{eq:approximate} achieves approximate state merging of $\Ket{\psi}^{RAB}$ within $\epsilon$.
  To calculate the error in approximate state merging,
  we use the purified distance $P\left(\rho,\sigma\right)$ of any two normalized states $\rho$ and $\sigma$ defined as
  \[
    P\left(\rho,\sigma\right)\coloneqq\sqrt{1-F^2\left(\rho,\sigma\right)}.
  \]
  The purified distance has the following properties:
  \begin{enumerate}
    \item $P\left(\rho,\sigma\right)\leqq P\left(\rho,\tau\right)+P\left(\tau,\sigma\right)$ (triangle inequality);
    \item $P\left(\mathcal{E}\left(\rho\right),\mathcal{E}\left(\sigma\right)\right)\leqq P\left(\rho,\sigma\right)$ (monotonicity),
  \end{enumerate}
  where $\rho$, $\sigma$, and $\tau$ are any state, and $\mathcal{E}$ is any CPTP map~\cite{T5}.
  Moreover, for any state $\rho$, $\sigma$, and $\tau$,
  \[
    P\left(\rho\otimes\tau,\sigma\otimes\tau\right)=P\left(\rho,\sigma\right),
  \]
  due to the multiplicativity of the fidelity~\cite{W5}.
  Using these properties, we obtain
  \begin{align*}
    &P\left(\id^R\otimes\tilde{\mathcal{M}}\left({\psi}^{RAB}\otimes{\Phi^+_K}^{\overline{A}\overline{B}}\right),{\psi}^{RB'B}\otimes{\Phi^+_L}^{\overline{A}\overline{B}}\right)\\
    &\leqq P\Big(\id^R\otimes\tilde{\mathcal{M}}\left({\psi}^{RAB}\otimes{\Phi^+_K}^{\overline{A}\overline{B}}\right),\\
    &\qquad\id^R\otimes\tilde{\mathcal{M}}\left({\tilde\psi}^{RAB}\otimes{\Phi^+_K}^{\overline{A}\overline{B}}\right)\Big)\\
    &\quad +P\Big(\id^R\otimes\tilde{\mathcal{M}}\left({\tilde\psi}^{RAB}\otimes{\Phi^+_K}^{\overline{A}\overline{B}}\right),\\
    &\qquad{\psi}^{RB'B}\otimes{\Phi^+_L}^{\overline{A}\overline{B}}\Big)\\
    &\leqq P\left({\psi}^{RAB}\otimes{\Phi^+_K}^{\overline{A}\overline{B}},{\tilde\psi}^{RAB}\otimes{\Phi^+_K}^{\overline{A}\overline{B}}\right)\\
    &\quad +P\left({\tilde\psi}^{RB'B}\otimes{\Phi^+_L}^{\overline{A}\overline{B}},{\psi}^{RB'B}\otimes{\Phi^+_L}^{\overline{A}\overline{B}}\right)\\
    &= P\left({\psi}^{RAB},{\tilde\psi}^{RAB}\right) +P\left({\tilde\psi}^{RB'B},{\psi}^{RB'B}\right)\\
    &\leqq\frac{\epsilon}{2}+\frac{\epsilon}{2}\\
    &=\epsilon.
  \end{align*}
  Therefore,
  \[
    \begin{split}
      &F^2\left(\id^R\otimes\tilde{\mathcal{M}}\left({\psi}^{RAB}\otimes{\Phi^+_K}^{\overline{A}\overline{B}}\right),{\psi}^{RB'B}\otimes{\Phi^+_L}^{\overline{A}\overline{B}}\right)\\
      &\geqq 1-\epsilon^2.
    \end{split}
  \]
\end{IEEEproof}

\subsection{\label{sec:converse_approximate}Improved converse bound for approximate state merging}
Given any pure state $\Ket{\psi}^{RAB}$ and an error $\epsilon\geqq 0$,
we extend Theorem~\ref{thm:new} and obtain a converse bound of entanglement cost in approximate state merging of $\Ket{\psi}^{RAB}$ within $\epsilon$.
We also show that our converse bound for approximate state merging improves the converse bound derived from the previous study on one-shot approximate state redistribution~\cite{B10}, when $\epsilon$ is sufficiently small.

In the same way as the proof of Theorem~\ref{thm:new} on exact state merging,
we obtain a converse bound of entanglement cost in approximate state merging by applying a majorization condition for LOCC convertibility to the bipartition between $B$ and $RA$.
In the following, for any hermitian operator $\rho$, we write a real vector of the eigenvalues of $\rho$ sorted in descending order as $\boldsymbol{\lambda}\left(\rho\right)$.
While the proof of Theorem~\ref{thm:new} on exact state merging uses the majorization condition for LOCC convertibility between bipartite pure states~\cite{N2},
approximate state merging requires another majorization condition for LOCC convertibility from a bipartite pure state to a bipartite mixed state, since the final state in approximate state merging can be a mixed state.
Reference~\cite{J1} provides a characterization of LOCC convertibility from a bipartite pure state to a bipartite mixed state as follows.

\begin{lemma}
\label{lem:mixed}
  (Theorem~1 in Ref.~\cite{J1})
  \textit{LOCC convertibility from a bipartite pure state to a bipartite mixed state.}
  Consider two spatially separated parties $X$ and $Y$ and systems $\mathcal{H}^X$ of $X$ and $\mathcal{H}^Y$ of $Y$.
  Any pure state $\Ket{\psi}^{XY}$ can be transformed into a mixed state $\rho^{XY}$ deterministically and exactly by LOCC if and only if
  \[
    \boldsymbol{\lambda}\left(\psi^Y\right)\prec\min\sum_j p(j)\boldsymbol{\lambda}\left(\psi_j^Y\right),
  \]
  where $\prec$ denotes majorization for real vectors~\cite{W11}, and the minimization is taken over any ensemble ${\left\{p(j),\Ket{\psi_j}^{XY}\right\}}_j$ of pure states which are not necessarily orthogonal to each other and satisfy $\rho^{XY}=\sum_j p(j)\Ket{\psi_j}\Bra{\psi_j}^{XY}$.
\end{lemma}

Given any pure state $\Ket{\psi}^{RAB}$ and an error $\epsilon\geqq 0$,
we obtain a converse bound of entanglement cost of approximate state merging of $\Ket{\psi}^{RAB}$ within $\epsilon$ using Lemma~\ref{lem:mixed} as follows.

\begin{theorem}
\label{thm:approximate_converse}
\textit{A converse bound of entanglement cost in approximate state merging.}
For any state $\Ket{\psi}^{RAB}$, any error $\epsilon\geqq 0$, and any algorithm for approximate state merging of $\Ket{\psi}^{RAB}$ within $\epsilon$,
it holds that
\[
  \begin{split}
    &\log_2 K - \log_2 L \geqq\\
    &\inf\Big\{\log_2 K - \log_2 L:\\
    &\boldsymbol{\lambda}\left(\psi^{B}\otimes\frac{\mathbbm{1}_K^{\overline{B}}}{K}\right)\prec\sum_j p(j)\boldsymbol{\lambda}\left(\psi_j^{B^\prime B\overline{B}}\right),\\
    &F^2\Big(\sum_j p(j)\Ket{\psi_j}\Bra{\psi_j}^{RB^\prime B\overline{A}\overline{B}},\psi^{RB^\prime B}\otimes{\Phi_L^+}^{\overline{A}\overline{B}}\Big)\geqq 1-\epsilon^2\Big\}.
  \end{split}
\]
\end{theorem}

\begin{IEEEproof}
  Any algorithm for approximate state merging transforms $\Ket{\psi}^{RAB}\otimes\Ket{\Phi_K^+}^{\overline{A}\overline{B}}$ into $\rho^{RB^\prime B \overline{A}\overline{B}}$ by LOCC, where $\rho$ satisfies
  \[
    F^2\left(\rho^{RB^\prime B \overline{A}\overline{B}},\psi^{RB^\prime B}\otimes{\Phi_L^+}^{\overline{A}\overline{B}}\right)\geqq 1-\epsilon^2.
  \]
  Substituting $X$, $Y$, and $\Ket{\psi}^{XY}$ in Lemma~\ref{lem:mixed} with $R\overline{A}$, $B^\prime B \overline{B}$, and $\Ket{\psi}^{RAB}\otimes\Ket{\Phi_K^+}^{\overline{A}\overline{B}}$, respectively, we obtain an ensemble ${\left\{p(j),\Ket{\psi_j}^{RB^\prime B \overline{A}\overline{B}}\right\}}_j$ satisfying
  \[
    \begin{split}
      &\rho^{RB^\prime B \overline{A}\overline{B}}=\sum_j p(j)\Ket{\psi_j}\Bra{\psi_j}^{RB^\prime B \overline{A}\overline{B}},\\
      &\boldsymbol{\lambda}\left(\psi^{B}\otimes\frac{\mathbbm{1}_K^{\overline{B}}}{K}\right)\prec\sum_j p(j)\boldsymbol{\lambda}\left(\psi_j^{B^\prime B\overline{B}}\right).
    \end{split}
  \]
  Therefore, we obtain the conclusion.
\end{IEEEproof}

Reference~\cite{B10} also analyzes a converse bound for fully quantum algorithms for one-shot approximate state redistribution, which is a generalized task including approximate state merging as a special case.
As discussed in Remark~\ref{remark:usefulness} in Sec.~\ref{sec:merge} in the main text, it is straightforward to convert this converse bound for fully quantum algorithms to the converse bound of entanglement cost in the LOCC framework, and we obtain the following lemma.

\begin{lemma}
\label{lem:existing_approxiamte_converse}
  (Proposition~12 in Ref.~\cite{B10})
  \textit{A converse bound of entanglement cost in approximate state merging shown in Ref.~\cite{B10}.}
  For any state $\Ket{\psi}^{RAB}$, any errors $\epsilon_1\in(0,1)$, $\epsilon_2\in(0,1-\epsilon_1)$, and any algorithm for approximate state merging of $\Ket{\psi}^{RAB}$ within $\epsilon_1$,
  it holds that
  \[
    \log_2 K - \log_2 L \geqq {H_{\min}^{\epsilon_2}(AB)}_\psi-{H_{\min}^{\epsilon_1+\epsilon_2}(B)}_\psi,
  \]
  where $H_{\min}^{\epsilon}$ is the smooth min-entropy~\cite{R2,T5}.
\end{lemma}

When the error tolerance in approximate state merging is sufficiently small,
our converse bound shown in Theorem~\ref{thm:approximate_converse} improves the converse bound shown in Lemma~\ref{lem:existing_approxiamte_converse} in the following sense.

\begin{proposition}
  \textit{Comparison of converse bounds of entanglement cost in approximate state merging.}
  For any state $\Ket{\psi}^{RAB}$, any errors $\epsilon_1\in(0,1)$, $\epsilon_2\in(0,1-\epsilon_1)$, and any algorithm for approximate state merging of $\Ket{\psi}^{RAB}$ within $\epsilon_1$,
  it holds that
  \[
    \begin{split}
      &\lim_{\epsilon_1\to 0}\inf\Big\{\log_2 K - \log_2 L:\\
      &\boldsymbol{\lambda}\left(\psi^{B}\otimes\frac{\mathbbm{1}_K^{\overline{B}}}{K}\right)\prec\sum_j p(j)\boldsymbol{\lambda}\left(\psi_j^{B^\prime B\overline{B}}\right),\\
      &F^2\Big(\sum_j p(j)\Ket{\psi_j}\Bra{\psi_j}^{RB^\prime B\overline{A}\overline{B}},\psi^{RB^\prime B}\otimes{\Phi_L^+}^{\overline{A}\overline{B}}\Big)\geqq 1-\epsilon_1^2\Big\}\\
      &\geqq\lim_{\epsilon_1,\epsilon_2\to 0}\left({H_{\min}^{\epsilon_2}(AB)}_\psi-{H_{\min}^{\epsilon_1+\epsilon_2}(B)}_\psi\right),
    \end{split}
  \]
  where $\epsilon_2\in(0,1-\epsilon_1)$ is arbitrary, and the notations are the same as those in Theorem~\ref{thm:approximate_converse} and Lemma~\ref{lem:existing_approxiamte_converse}.
\end{proposition}

\begin{IEEEproof}
  Regarding our converse bound, it holds that
  \[
    \begin{split}
      &\lim_{\epsilon_1\to 0}\inf\Big\{\log_2 K - \log_2 L:\\
      &\boldsymbol{\lambda}\left(\psi^{B}\otimes\frac{\mathbbm{1}_K^{\overline{B}}}{K}\right)\prec\sum_j p(j)\boldsymbol{\lambda}\left(\psi_j^{B^\prime B\overline{B}}\right),\\
      &F^2\Big(\sum_j p(j)\Ket{\psi_j}\Bra{\psi_j}^{RB^\prime B\overline{A}\overline{B}},\psi^{RB^\prime B}\otimes{\Phi_L^+}^{\overline{A}\overline{B}}\Big)\geqq 1-\epsilon_1^2\Big\}\\
      &=\inf\left\{\log_2 K - \log_2 L: \frac{\mathbbm{1}_K}{K}\otimes\psi^{B}\prec\frac{\mathbbm{1}_L}{L}\otimes\psi^{AB}\right\}.
    \end{split}
  \]
  As for the converse bound shown in Lemma~\ref{lem:existing_approxiamte_converse}, it holds that
  \[
    \lim_{\epsilon_1,\epsilon_2\to 0}\left({H_{\min}^{\epsilon_2}(AB)}_\psi-{H_{\min}^{\epsilon_1+\epsilon_2}(B)}_\psi\right)=\log_2 \frac{1}{\lambda_0^{AB}} - \log_2 \frac{1}{\lambda_0^{B}},
  \]
  where $\lambda_0^{AB}$ and $\lambda_0^{B}$ are the largest eigenvalues of $\psi^{AB}$ and $\psi^B$, respectively.

  The majorization
  \[
    \frac{\mathbbm{1}_K}{K}\otimes\psi^{B}\prec\frac{\mathbbm{1}_L}{L}\otimes\psi^{AB}
  \]
  implies that the largest eigenvalues of this majorization satisfy
  \[
    \frac{\lambda_0^{B}}{K}\leqq\frac{\lambda_0^{AB}}{L},
  \]
  and hence,
  \[
    \log_2 K - \log_2 L \geqq\log_2 \frac{1}{\lambda_0^{AB}} - \log_2 \frac{1}{\lambda_0^{B}}.
  \]
  Due to this implication, we obtain
  \[
    \begin{split}
      &\inf\left\{\log_2 K - \log_2 L: \frac{\mathbbm{1}_K}{K}\otimes\psi^{B}\prec\frac{\mathbbm{1}_L}{L}\otimes\psi^{AB}\right\}\\
      &\geqq\log_2 \frac{1}{\lambda_0^{AB}} - \log_2 \frac{1}{\lambda_0^{B}},
    \end{split}
  \]
  which yields the conclusion.
\end{IEEEproof}

\section{\label{sec:split}Exact state splitting}

In this appendix, we analyze entanglement cost of exact state splitting, which is an inverse task of exact state merging.
After giving the definition in Sec.~\ref{sec:def_split}, we proceed to provide the results in Sec.~\ref{sec:split_result}.

\subsection{\label{sec:def_split}Definition of exact state splitting}
Exact state splitting is an inverse task of exact state merging involving three parties $A$, $B$ and $R$,
where $R$ is a reference to consider purification.
By convention, for exact state splitting, we assign $A$ as a sender and $B$ as a receiver.
Let $A$ have systems $\mathcal{H}^A$, $\mathcal{H}^{A^\prime }$, and $\mathcal{H}^{\overline{A}}$, $B$ have $\mathcal{H}^B$ and $\mathcal{H}^{\overline{B}}$, and $R$ have $\mathcal{H}^R$,
where $\dim\mathcal{H}^{A^\prime }=\dim\mathcal{H}^B$.
We assume that $A$ and $B$ can freely perform local operations and classical communication (LOCC) assisted by a maximally entangled resource state initially shared between $\mathcal{H}^{\overline{A}}$ and $\mathcal{H}^{\overline{B}}$.
We write the maximally entangled resource state as
\[
  \Ket{\Phi^+_K}^{\overline{A} \overline{B}}\coloneqq\frac{1}{\sqrt{K}}\sum_{l=0}^{K-1}\Ket{l}^{\overline{A}} \otimes\Ket{l}^{\overline{B}},
\]
where $K$ denotes the Schmidt rank of the resource state.
Note that $A$ and $B$ cannot perform any operation on $\mathcal{H}^R$.

\begin{figure}[t]
  \centering
  \includegraphics[width=3.5in]{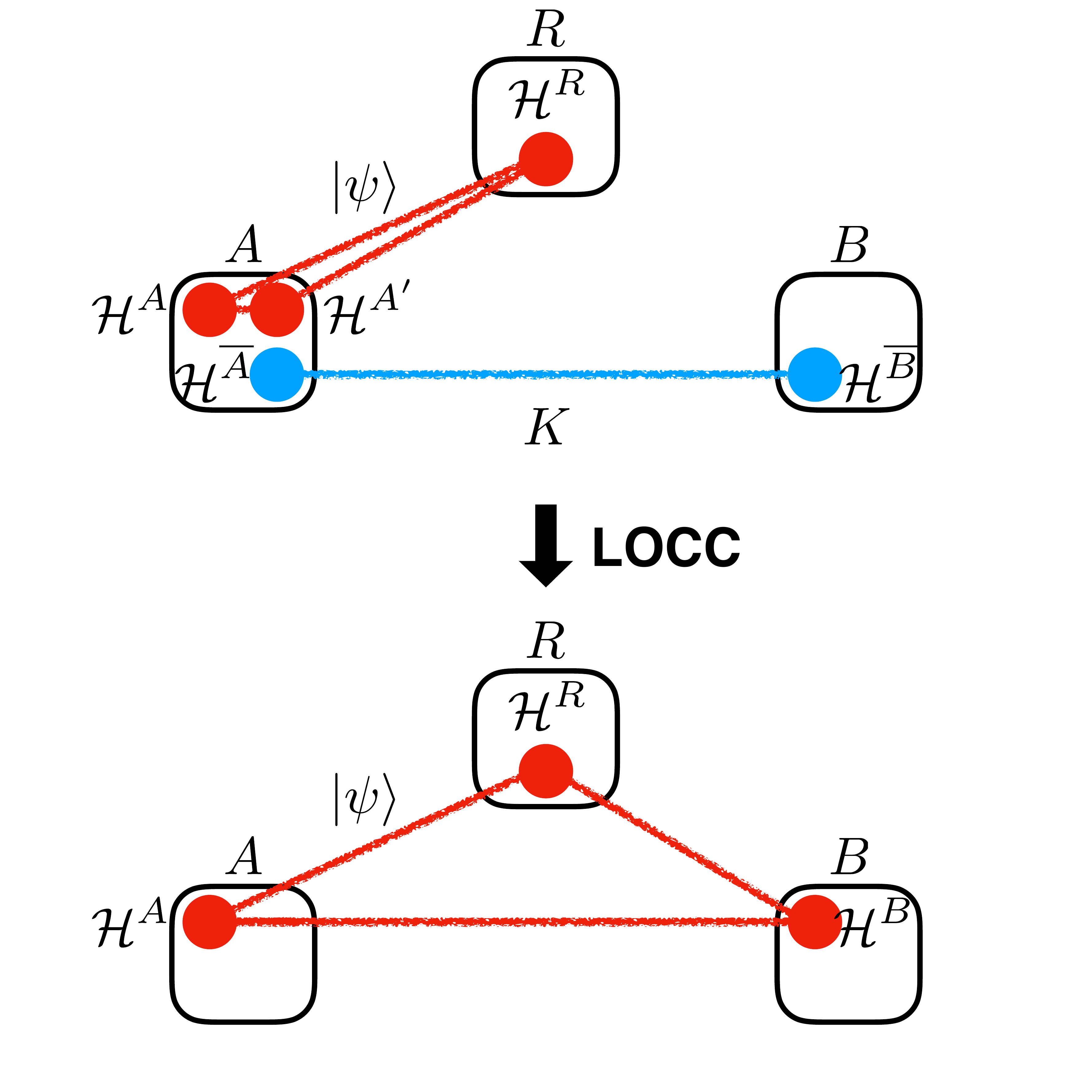}
  \caption{\label{fig:split}Exact state splitting of a given state $\Ket{\psi}^{RAA^\prime }$ denoted by the red circles.   Parties $A$ and $B$ perform LOCC assisted by a maximally entanglement resource state $\Ket{\Phi_K^+}^{\overline{AB}}$ with the Schmidt rank $K$ denoted by the blue circles to transfer the reduced state $\psi^{A^\prime }$ from $A$ to $B$ and obtain $\Ket{\psi}^{RAB}$. }
\end{figure}

We define the task of exact state splitting as illustrated in Fig.~\ref{fig:split}.
Initially, $A$ is given a possibly mixed state of $\mathcal{H}^A\otimes\mathcal{H}^{A^\prime}$ whose purification is represented by $\Ket{\psi}^{R A A^\prime }$, where $A$ and $B$ knows classical description of $\Ket{\psi}^{R A A^\prime }$.
Exact state splitting of $\Ket{\psi}^{R A A^\prime }$ is a task for $A$ and $B$ to transfer a part of $\Ket{\psi}^{R A A^\prime }$ corresponding to $\mathcal{H}^{A^\prime}$ from $A$ to $B$ and obtain $\Ket{\psi}^{R A B}$.

\begin{definition}
\label{def:splitting}
    \textit{Exact state splitting.}
    Exact state splitting of a purified given state $\Ket{\psi}^{R A A^\prime }$ is a task for parties $A$ and $B$ to achieve a transformation
      \[
        \begin{split}
            \id^R \otimes\mathcal{S}\left({\psi}^{RAA^\prime }\otimes{\Phi^+_K}^{\overline{A}\overline{B}}\right)
            ={\psi}^{RAB}
        \end{split}
    \]
    by an LOCC map $\mathcal{S}: \mathcal{B}\left(\mathcal{H}^A\otimes\mathcal{H}^{A^\prime }\otimes\mathcal{H}^{\overline{A}}\otimes\mathcal{H}^{\overline{B}}\right) \to \mathcal{B}\left(\mathcal{H}^A\otimes\mathcal{H}^B\otimes\mathcal{H}^{\overline{A}}\otimes\mathcal{H}^{\overline{B}}\right)$, which can be constructed depending on the classical description of $\Ket{\psi}^{RAA^\prime}$.
\end{definition}

Entanglement cost of exact state splitting is defined as $\log_2 K$.
If $\log_2 K\geqq\log_2 \dim\mathcal{H}^{A^\prime}$, there exists a trivial algorithm for exact state splitting by quantum teleportation to transfer $\psi^{A^\prime }$ from $A$ to $B$.
In contrast, our algorithm presented in Sec.~\ref{sec:split_result} exploits classical description of $\Ket{\psi}^{RAA^\prime }$ for reducing entanglement cost, and it is shown to be an optimal algorithm achieving the minimal entanglement cost.

\subsection{\label{sec:split_result}Optimal algorithm for exact state splitting}

We derive a formula for the minimal entanglement cost of an algorithm for exact state splitting.
For exact state splitting of $\Ket{\psi}^{RAA^\prime }$, the following theorem yields the minimal entanglement cost and an optimal algorithm.

\begin{theorem}
\label{thm:split}
    \textit{Optimal entanglement cost of exact state splitting.}
    Given any pure state $\Ket{\psi}^{RAA^\prime }$,
    exact state splitting of $\Ket{\psi}^{RAA^\prime }$ is achievable if and only if
    \[
        \log_2 K \geqq \log_2 \rank \psi^{A^\prime }.
    \]
\end{theorem}

\begin{IEEEproof}
    \textit{If part}:
    We construct an LOCC algorithm achieving
    \begin{equation}
        \label{eq:split_upper}
        \log_2 K = \log_2 \rank \psi^{A^\prime }.
    \end{equation}
    Note that the trivial algorithm, that is, quantum teleportation of $\psi^{A^\prime }$, requires entanglement cost $\log_2 K = \log_2\dim\mathcal{H}^{A^\prime }$,
    and our algorithm achieving Eq.~\eqref{eq:split_upper} outperforms this trivial algorithm when $\psi^{A^\prime}$ is not a full-rank state, that is, $\rank\psi^{A^\prime }<\dim\mathcal{H}^{A^\prime}$; \textit{e.g.},
    when $\psi^{A^\prime}$ is locally represented as a code state of a quantum error correcting code using a larger-dimensional system $\mathcal{H}^{A^\prime}$ than the rank of $\psi^{A^\prime}$.

    To achieve Eq.~\eqref{eq:split_upper}, we provide a method for compressing $\psi^{A^\prime }$.
    Consider the Schmidt decomposition of the given state $\Ket{\psi}^{RAA^\prime }$ with respect to the bipartition between $\mathcal{H}^R\otimes\mathcal{H}^A$ and $\mathcal{H}^{A^\prime }$, that is,
    \[
        \Ket{\psi}^{RAA^\prime }=\sum_{l\in R_{\psi}} \sqrt{\lambda_l^\psi}\Ket{l}^{RA}\otimes\Ket{l}^{A^\prime },
    \]
    where $R_{\psi}\coloneqq\left\{0,\ldots,\rank \psi^{A^\prime }-1\right\}$, each $\sqrt{\lambda_l^\psi}>0$ is a nonzero Schmidt coefficient, and ${\left\{\Ket{l}^{RA}: l\in R_{\psi}\right\}}$ and ${\left\{\Ket{l}^{A^\prime }: l\in R_{\psi}\right\}}$ are subsets of the Schmidt bases of $\mathcal{H}^R\otimes\mathcal{H}^A$ and $\mathcal{H}^{A^\prime }$, respectively, corresponding to the nonzero Schmidt coefficients.
    Let $\mathcal{H}^{A^{\prime\prime}}$ be $A$'s auxiliary system satisfying $\dim\mathcal{H}^{A^{\prime\prime}}=\rank \psi^{A^\prime }$
    and ${\left\{\Ket{l}^{A^{\prime\prime}}: l\in R_{\psi}\right\}}$ be the computational basis of $\mathcal{H}^{A^{\prime\prime}}$.
    Consider an isometry $U_\textup{split}$ from $\mathcal{H}^{A^\prime }$ to $\mathcal{H}^{A^{\prime\prime}}$ satisfying $\Ket{l}^{A^{\prime\prime}}=U_\textup{split}\Ket{l}^{A^\prime }$
    for each $l\in R_{\psi}$.
    By performing $U_\textup{split}$, $\psi^{A^\prime }$ is compressed into a state on $\mathcal{H}^{A^{\prime\prime}}$, that is,
    \[
        \begin{split}
          \Ket{\psi'}^{RAA^{\prime\prime}}&\coloneqq\mathbbm{1}^{RA}\otimes U_\textup{split}\Ket{\psi}^{RAA^\prime }\\
                             &=\sum_{l\in R_{\psi}} \sqrt{\lambda_l^\psi}\Ket{l}^{RA}\otimes\Ket{l}^{A^{\prime\prime}}.
        \end{split}
    \]
    By performing $U_\textup{split}^\dag$, the given state $\Ket{\psi}$ can be recovered from the compressed state $\Ket{\psi'}$.

    The LOCC algorithm achieving Eq.~\eqref{eq:split_upper} is as follows.
    First, $A$ performs $U_\textup{split}$ to transform the given state $\Ket{\psi}^{RAA^\prime }$ into the compressed state $\Ket{\psi'}^{RAA^{\prime\prime}}$.
    Next, the reduced state ${\psi'}^{A^{\prime\prime}}$ is sent from $A$ to $B$ by quantum teleportation using the resource state satisfying Eq.~\eqref{eq:split_upper}.
    After performing quantum teleportation, $B$ performs $U_\textup{split}^\dag$ on the system for the received state to recover $\Ket{\psi}^{RAB}$.

    \textit{Only if part}:
    We use LOCC monotonicity of the Schmidt rank~\cite{L2}.
    The Schmidt rank of $\Ket{\psi}^{RAA^\prime }\otimes\Ket{\Phi^+_K}^{\overline{A}\overline{B}}$ between the party $B$ and the other parties $R$ and $A$ is $K$.
    After performing an LOCC map $\id^R\otimes\mathcal{S}$, the Schmidt rank of $\Ket{\psi}^{RAB}$ between the party $B$ and the other parties $R$ and $A$ is $\rank\psi^{A^\prime }$.
    Since the Schmidt rank of pure states is monotonically non-increasing under LOCC,
    it holds that $K\geqq\rank\psi^{A^\prime }$.
    Therefore, we obtain $\log_2 K \geqq \log_2 \rank \psi^{A^\prime }$.
\end{IEEEproof}

\begin{remark}
    \textit{Asymptotic limit of exact state splitting.}
    Given any pure state $\Ket{\psi}^{RAA^\prime }$,
    from our exact algorithm for one-shot state splitting in Theorem~\ref{thm:split},
    we can derive the rate of entanglement cost required for \textit{asymptotic} state splitting of $\Ket\psi$ in the LOCC framework as follows.
    Note that the asymptotic rate derived in the following, that is, ${H\left(A^\prime\right)}_\psi$, is optimal~\cite{D2}, where $H$ denotes the quantum entropy~\cite{W5}.

    For large $n$, $\Ket{\psi}^{\otimes n}$ can be approximated by $\ket{\tilde{\psi}^n}\coloneqq\left({\left(\mathbbm{1}^{RA}\right)}^{\otimes n}\otimes\Pi_{\delta}^{{A^\prime }^n}\right)\Ket{\psi}^{\otimes n}\approx\Ket{\psi}^{\otimes n}$,
    where $\Pi_{\delta}^{{A^\prime }^n}$ is the projector onto the $\delta$-typical subspace of ${\left(\psi^{A^\prime }\right)}^{\otimes n}$~\cite{W5},
    and ${\left(\mathbbm{1}^{RA}\right)}^{\otimes n}$ is the identity operator on ${\left(\mathcal{H}^R\otimes\mathcal{H}^A\right)}^{\otimes n}$.
    Then, entanglement cost $\log_2 K$ of exact state splitting of $\ket{\tilde{\psi}^n}$ yields ${H\left(A^\prime \right)}_\psi$ required for the asymptotic state splitting of $\Ket\psi$ because
    \begin{align*}
      &\frac{1}{n}\log_2 K\\
      &=\frac{1}{n}\log_2\rank\tr_{RA}\ket{\tilde{\psi}^n}\bra{\tilde{\psi}^n}\\
      &\leqq\frac{1}{n}\log_2\rank\Pi_{\delta}^{{A^\prime }^n}\\
      &={H\left(A^\prime \right)}_\psi+o\left(\frac{1}{n}\right)\quad \text{as $n\to\infty$}.
    \end{align*}
\end{remark}

\section{\label{sec:equivalence}Tasks equivalent to exact state merging}

In this appendix, we show the following equivalent tasks of exact state merging of a given state $\Ket{\psi}^{RAB}$,
in the sense that these tasks are achievable at the same entanglement cost using the same algorithm.

\begin{proposition}
\label{prp:max}
  \textit{Equivalence of exact state merging of an arbitrary tripartite pure state, a corresponding maximally entangled state, and a corresponding set of bipartite states.}
  Given any fixed integer $K$, $L$, and any pure state $\Ket{\psi}^{RAB}$ whose Schmidt rank with respect to bipartition between $\mathcal{H}^R$ and $\mathcal{H}^{A}\otimes\mathcal{H}^{B}$ is $D$ and whose Schmidt decomposition is given by Eq.~\eqref{eq:schmidt} in the main text,
  the following statements are equivalent:
  \begin{enumerate}
    \item An LOCC map $\mathcal{M}$ achieves the following exact state merging of $\Ket{\psi}^{RAB}$
      \[
        \begin{split}
          \id^R\otimes\mathcal{M}\left({\psi}^{RAB}\otimes{\Phi^+_K}^{\overline{A}\overline{B}}\right)
          ={\psi}^{RB^\prime B}\otimes{\Phi^+_L}^{\overline{A}\overline{B}};
        \end{split}
      \]
    \item The same LOCC map $\mathcal{M}$ as the above achieves the following exact state merging of $\Ket{\Phi_D^+\left(\psi\right)}^{RAB}$
      \[
        \begin{split}
          &\id^R\otimes\mathcal{M}\left({\Phi_D^+\left(\psi\right)}^{RAB}\otimes{\Phi^+_K}^{\overline{A}\overline{B}}\right)\\
          &={\Phi_D^+\left(\psi\right)}^{RB'B}\otimes{\Phi^+_L}^{\overline{A}\overline{B}},
        \end{split}
      \]
      where $\Ket{\Phi_D^+\left(\psi\right)}^{RAB}$ is the maximally entangled state corresponding to $\Ket{\psi}^{RAB}$, defined as Eq.~\eqref{eq:max} in the main text.
    \item Define a set $S_\psi^{AB}\subset\mathcal{D}\left(\mathcal{H}^A\otimes\mathcal{H}^B\right)$ of arbitrary bipartite states on a subspace of $\mathcal{H}^A\otimes\mathcal{H}^B$ spanned by the Schmidt-basis states ${\left\{\Ket{\psi_l}^{AB}\right\}}_l$ corresponding to nonzero Schmidt coefficients of $\Ket{\psi}^{RAB}$ in Eq.~\eqref{eq:schmidt}, that is,
      \[
        \begin{split}
          S_\psi^{AB}&\coloneqq{\left\{\psi_{\boldsymbol{\alpha}}^{AB}\in\mathcal{D}\left(\mathcal{H}^A\otimes\mathcal{H}^B\right)\right\}}_{\boldsymbol{\alpha}},\\
          \psi_{\boldsymbol{\alpha}}^{AB}&\coloneqq\sum_{l=0}^{D-1}\sum_{l^\prime=0}^{D-1}\alpha_{l,l^\prime} \Ket{\psi_l}\Bra{\psi_{l^\prime}}^{AB},
        \end{split}
      \]
      where $\boldsymbol{\alpha}$ denotes the tuple of the parameters $\alpha_{l,l^\prime}$ for all $l$ and $l^\prime$.
      Then, the same LOCC map $\mathcal{M}$ as the above achieves the following state transformation for any bipartite state $\psi_{\boldsymbol{\alpha}}^{AB}\in S_\psi^{AB}$
      \[
        \begin{split}
          \mathcal{M}\left(\psi_{\boldsymbol{\alpha}}^{AB}\otimes{\Phi^+_K}^{\overline{A}\overline{B}}\right)
          =\psi_{\boldsymbol{\alpha}}^{B^\prime B}\otimes{\Phi^+_L}^{\overline{A}\overline{B}},
        \end{split}
      \]
      where $\mathcal{M}$ is independent of $\boldsymbol{\alpha}$.
  \end{enumerate}
  The same equivalence also holds for non-catalytic exact state merging if we fix $\log_2 L=0$.
\end{proposition}

\begin{IEEEproof}
  We prove the equivalence in the catalytic setting while the statement on non-catalytic exact state merging follows from the same argument setting $\log_2 L=0$.
  We show that each of Statements~1--3 holds if and only if
  \begin{equation}
    \label{eq:m}
    \mathcal{M}\left({\Ket{\psi_l}\Bra{\psi_{l'}}}^{AB}\otimes{\Phi^+_K}^{\overline{A}\overline{B}}\right)={\Ket{\psi_l}\Bra{\psi_{l'}}}^{B'B}\otimes{\Phi^+_L}^{\overline{A}\overline{B}}
  \end{equation}
  holds for any $l$ and $l'$.

  \textit{Statement~1 $\Leftrightarrow$ \textup{Eq.}~\eqref{eq:m}}:
  Assume Statement~1; that is, an LOCC map $\mathcal{M}$ by $A$ and $B$ achieves the following exact state merging of $\Ket{\psi}^{RAB}$
  \[
    \id^R\otimes\mathcal{M}\left({\psi}^{RAB}\otimes{\Phi^+_K}^{\overline{A}\overline{B}}\right)={\psi}^{RB'B}\otimes{\Phi^+_L}^{\overline{A}\overline{B}}.
  \]
  The left-hand side and the right-hand side are written as
  \[
    \begin{split}
      &\id^R\otimes\mathcal{M}\left({\psi}^{RAB}\otimes{\Phi^+_K}^{\overline{A}\overline{B}}\right)\\
      &=\sum_{l,l'}\frac{1}{\sqrt{\lambda_l\lambda_{l'}}}{\Ket{l}\Bra{l'}}^R\otimes\mathcal{M}\left({\Ket{\psi_l}\Bra{\psi_{l'}}}^{AB}\otimes{\Phi^+_K}^{\overline{A}\overline{B}}\right),\\
      &{\psi}^{RB'B}\otimes{\Phi^+_L}^{\overline{A}\overline{B}}\\
      &=\sum_{l,l'}\frac{1}{\sqrt{\lambda_l\lambda_{l'}}}{\Ket{l}\Bra{l'}}^R\otimes{\Ket{\psi_l}\Bra{\psi_{l'}}}^{B'B}\otimes{\Phi^+_L}^{\overline{A}\overline{B}}.
    \end{split}
  \]
  Due to the linear independence, we obtain Eq.~\eqref{eq:m} for any $l$ and $l'$.
  The converse follows from the linearity of $\mathcal{M}$.

  \textit{Statement~2 $\Leftrightarrow$ \textup{Eq.}~\eqref{eq:m}}: This equivalence can be shown in the same way as the equivalence between Statement~1 and Eq.~\eqref{eq:m}, by substituting $\psi$ with ${\Phi}_D^+\left(\psi\right)$.

  \textit{Statement~3 $\Leftrightarrow$ \textup{Eq.}~\eqref{eq:m}}: Assume Statement~3. For each $l$,
  \[
    \mathcal{M}\left({\Ket{\psi_l}\Bra{\psi_{l}}}^{AB}\otimes{\Phi^+_K}^{\overline{A}\overline{B}}\right)={\Ket{\psi_l}\Bra{\psi_{l}}}^{B'B}\otimes{\Phi^+_L}^{\overline{A}\overline{B}}
  \]
  holds as a special case of Statement~3.
  For any different $l$ and $l^\prime$,
  consider two cases of choosing $\psi_{\boldsymbol{\alpha}}^{AB}\in S_\psi^{AB}$ as
  \[
    \begin{split}
      \frac{1}{2}\Ket{\psi_l}\Bra{\psi_l}+\frac{1}{2}\Ket{\psi_l}\Bra{\psi_{l^\prime}}+\frac{1}{2}\Ket{\psi_{l^\prime}}\Bra{\psi_l}+\frac{1}{2}\Ket{\psi_{l^\prime}}\Bra{\psi_{l^\prime}}
    \end{split}
  \]
  and
  \[
    \begin{split}
      \frac{1}{2}\Ket{\psi_l}\Bra{\psi_l}+\frac{\textup{i}}{2}\Ket{\psi_l}\Bra{\psi_{l^\prime}}-\frac{\textup{i}}{2}\Ket{\psi_{l^\prime}}\Bra{\psi_l}+\frac{1}{2}\Ket{\psi_{l^\prime}}\Bra{\psi_{l^\prime}}.
    \end{split}
  \]
  Applying Statement~3 to these two states and using the linearity of $\mathcal{M}$, we obtain
  \[
    \begin{split}
      \mathcal{M}\left({\Ket{\psi_l}\Bra{\psi_{l^\prime}}}^{AB}\otimes{\Phi^+_K}^{\overline{A}\overline{B}}\right)&={\Ket{\psi_l}\Bra{\psi_{l^\prime}}}^{B'B}\otimes{\Phi^+_L}^{\overline{A}\overline{B}},\\
      \mathcal{M}\left({\Ket{\psi_{l^\prime}}\Bra{\psi_{l}}}^{AB}\otimes{\Phi^+_K}^{\overline{A}\overline{B}}\right)&={\Ket{\psi_{l^\prime}}\Bra{\psi_{l}}}^{B'B}\otimes{\Phi^+_L}^{\overline{A}\overline{B}}.
    \end{split}
  \]
  Therefore, Eq.~\eqref{eq:m} holds for any $l$ and $l^\prime$.
  The converse follows from the linearity of $\mathcal{M}$.
\end{IEEEproof}

\section{\label{sec:koashi_imoto_decomposition_algorithm}How to obtain Koashi-Imoto decomposition}

In this appendix, we demonstrate how to obtain the Koashi-Imoto decomposition of a given tripartite pure state required for the algorithms for exact state merging in Theorems~\ref{thm:merge} and~\ref{thm:merge_without_catalyst} in the main text.
The Koashi-Imoto decomposition of a tripartite pure state $\Ket{\psi}^{RAB}$ shown in Lemma~\ref{lem:koashi_imoto_decomposition_tripartite} in the main text follows from a decomposition defined for a corresponding set of states.
In Sec.~\ref{sec:tripartite_from_set}, we summarize the Koashi-Imoto decomposition of a given set of states and discuss how to obtain the Koashi-Imoto decomposition of $\Ket{\psi}^{RAB}$ from that of a set of states.
In Sec.~\ref{sec:decomposition_set}, we summarize an algorithm shown in Ref.~\cite{K3} for obtaining the Koashi-Imoto decomposition of any given set of states in terms of our notations, and provide an example of how to obtain the Koashi-Imoto decomposition of a given tripartite pure state using this algorithm.

\subsection{\label{sec:tripartite_from_set}Koashi-Imoto decomposition of a tripartite pure state from that of a set of states}

Given any tripartite pure state $\Ket{\psi}^{RAB}$,
we present how to obtain the Koashi-Imoto decomposition of $\Ket{\psi}^{RAB}$ from that of the corresponding set.
The Koashi-Imoto decomposition of a given set of states $\left\{\psi_i^A\in\mathcal{D}\left(\mathcal{H}^A\right):i\in I\right\}$ characterizes a CPTP map $\mathcal{T}:\mathcal{B}\left(\mathcal{H}^A\right)\to\mathcal{B}\left(\mathcal{H}^A\right)$ leaving any state in the set invariant.
Note that the index set $I$ can be an infinite set.
The Koashi-Imoto decomposition of a set of states is shown in the following lemma, of which an algorithmic proof is given in Ref.~\cite{K3}, and alternative proofs are given in Refs.~\cite{H6,K5} through an operator-algebraic approach.

\begin{lemma}
\label{lem:koashi_imoto_decomposition_set}
(Theorem~3 in Ref.~\cite{K3}, Theorem~9 in Ref.~\cite{H6}, and Lemma~6 in Ref.~\cite{K5})
    \textit{Koashi-Imoto decomposition of a set of states.}
    Given any set ${\left\{\psi_i^A\in\mathcal{D}\left(\mathcal{H}^A\right): i\in I\right\}}$,
    there exists a \textit{unique} decomposition of $\mathcal{H}^A$
    \[
        \mathcal{H}^A=\bigoplus_{j=0}^{J-1}\mathcal{H}^{a_j^L}\otimes\mathcal{H}^{a_j^R}
    \]
    such that
    \begin{enumerate}
        \item For each $i\in I$, $\psi_i^A$ is decomposed into
            \[
                \psi_i^A=\bigoplus_{j=0}^{J-1} p\left(j\right) \omega_j^{a_j^L}\otimes\phi_{i,j}^{a_j^R},
            \]
            where $p\left(j\right)$ is a probability distribution and for each $j\in\{0,\ldots,J-1\}$, $\omega_j^{a_j^L}\in\mathcal{D}\left(\mathcal{H}^{a_j^L}\right)$ is independent of $i$, and $\phi_{i,j}^{a_j^R}\in\mathcal{D}\left(\mathcal{H}^{a_j^R}\right)$ depends on $i$.
        \item For any CPTP map $\mathcal{T}:\mathcal{B}\left(\mathcal{H}^A\right)\to\mathcal{B}\left(\mathcal{H}^A\right)$,
            if $\mathcal{T}$ leaves $\psi_i^A$ invariant for each $i\in I$, that is, $\mathcal{T}\left(\psi_i^A\right)=\psi_i^A$,
            then any isometry $U_\mathcal{T}$ from $\mathcal{H}^A$ to $\mathcal{H}^A\otimes\mathcal{H}^{E}$ for $\mathcal{T}$'s Stinespring dilation $\mathcal{T}\left(\rho\right)=\tr_E U_\mathcal{T}\rho U_\mathcal{T}^\dag$ is decomposed into $U_\mathcal{T}=\bigoplus_{j=0}^{J-1} U_j^{a_j^L}\otimes\mathbbm{1}^{a_j^R}$,
            where, for each $j\in\{0,\ldots,J-1\}$, $U_j^{a_j^L}$ is an isometry from $\mathcal{H}^{a_j^L}$ to $\mathcal{H}^{a_j^L}\otimes\mathcal{H}^{E}$ satisfying $\tr_{E}U_\mathcal{T} \omega_j^{a_j^L} U_\mathcal{T}^\dag = \omega_j^{a_j^L}$.
    \end{enumerate}
\end{lemma}

Using Lemma~\ref{lem:koashi_imoto_decomposition_set}, Ref.~\cite{H6} considers the Koashi-Imoto decomposition of a given bipartite state $\psi^{RA}$.
Given any bipartite state $\psi^{RA}$, consider a set of states $S_\psi^{A|R}\coloneqq\left\{\psi^A\left(\Lambda^R\right):\Lambda^R\geqq 0\right\}$ defined as Eq.~\eqref{eq:psi_lambda} in the main text,
where we regard the operator $\Lambda^R$ as the index of the set.
Applying the Koashi-Imoto decomposition of a set of states shown in Lemma~\ref{lem:koashi_imoto_decomposition_set} to this set $S_\psi^{A|R}$,
we obtain the Koashi-Imoto decomposition of the bipartite state $\psi^{RA}$ as shown in the following lemma.

\begin{lemma}
\label{lem:koashi_imoto_decomposition_bipartite}
    (in Proof of Theorem~6 in Ref.~\cite{H6})
    \textit{Koashi-Imoto decomposition of a bipartite state.}
    Given any bipartite state $\psi^{RA}$,
    the Koashi-Imoto decomposition of the set $S_\psi^{A|R}$ defined as Eq.~\eqref{eq:psi_lambda} yields a unique decomposition of $\mathcal{H}^A$ satisfying the conditions in Lemma~\ref{lem:koashi_imoto_decomposition_set}
    \[
        \mathcal{H}^A=\bigoplus_{j=0}^{J-1}\mathcal{H}^{a_j^L}\otimes\mathcal{H}^{a_j^R},
    \]
    and $\psi^{RA}$ is decomposed into
    \[
        \psi^{RA}=\bigoplus_{j=0}^{J-1} p\left(j\right) \omega_j^{a_j^L}\otimes\phi_j^{Ra_j^R},
    \]
    where $p\left(j\right)$ is a probability distribution.
\end{lemma}

By considering a purification $\Ket{\psi}^{RAB}$ of the bipartite state $\psi^{RA}$ in Lemma~\ref{lem:koashi_imoto_decomposition_bipartite},
we obtain the Koashi-Imoto decomposition of a tripartite pure state $\Ket{\psi}^{RAB}$ shown in Lemma~\ref{lem:koashi_imoto_decomposition_tripartite} in the main text.
Consequently, to obtain the Koashi-Imoto decomposition of a given tripartite pure state $\Ket{\psi}^{RAB}$,
first apply the algorithm presented in Ref.~\cite{K3}, or the operator-algebraic theorems used in Refs.~\cite{H6,K5},
to the set of states $S_\psi^{A|R}$ defined as Eq.~\eqref{eq:psi_lambda},
and then follow the above argument.
The former way of applying the algorithm in Ref.~\cite{K3} is demonstrated in the next subsection of this appendix.

\subsection{\label{sec:decomposition_set}Algorithm for obtaining Koashi-Imoto decomposition}

We demonstrate how to obtain the Koashi-Imoto decomposition of a given tripartite pure state, using the algorithm presented in Ref.~\cite{K3} for obtaining the Koashi-Imoto decomposition of a set of states.

The algorithm shown in Ref.~\cite{K3} works by iteratively refining decompositions of the Hilbert space $\mathcal{H}^A$ in the form of
\begin{equation}
  \label{eq:decomposition_form}
  \mathcal{H}^A=\bigoplus_{j=0}^{J-1}\mathcal{H}^{a_j^L}\otimes\mathcal{H}^{a_j^R}.
\end{equation}
For a decomposition in this form, we let $\Pi^{a_j^L}$ and $\Pi^{a_j^R}$ denote the projectors onto $\mathcal{H}^{a_j^L}$ and $\mathcal{H}^{a_j^R}$, respectively.
The degree of refinement is evaluated by an index $r$ defined for the decomposition in the form of Eq.~\eqref{eq:decomposition_form} as
\[
  r\coloneqq\frac{1}{2}\left(\sum_{J=0}^{J-1}\dim\mathcal{H}^{a_j^R}\right)\left(\sum_{J=0}^{J-1}\dim\mathcal{H}^{a_j^R}+1\right)-J+1.
\]
The algorithm begins with initially regarding $\mathcal{H}^A$ as
\[
  \mathcal{H}^A=\mathcal{H}^{a_0^L},
\]
where $J=1$, the index is initially
\[
  r=1,
\]
and $\mathcal{H}^{a_0^R}$ does not explicitly appear since
\[
  \dim\mathcal{H}^{a_0^L}=\dim\mathcal{H}^A,\quad\dim\mathcal{H}^{a_0^R}=1.
\]
Then, the refinement can be performed by two types of procedures, which we name the \textit{$L$-decomposing procedure} and the \textit{$R$-combining procedure}.
According to the given set of states, the $L$-decomposing procedure decomposes a Hilbert space $\mathcal{H}^{a_{j_0}^L}$ in an intermediate decomposition in the form of Eq.~\eqref{eq:decomposition_form} into two subspaces, and the $R$-combining procedure combines two different Hilbert spaces $\mathcal{H}^{a_{j_0}^R}$ and $\mathcal{H}^{a_{j_1}^R}$ in an intermediate decomposition in the form of Eq.~\eqref{eq:decomposition_form} into one, as discussed later.
Each procedure increases the index $r$ representing the degree of refinement of the decomposition,
and the algorithm repeatedly applies either of the two procedures, until a decomposition maximizing $r$ is obtained.
Since $r$ is an integer bounded by
\[
  1\leqq r\leqq\frac{1}{2}\left(\dim\mathcal{H}^A\right)\left(\dim\mathcal{H}^A+1\right),
\]
the algorithm terminates after applying these procedures
\[
  O\left({\left(\dim\mathcal{H}^A\right)}^2\right)
\]
times in total.
The decomposition maximizing $r$ is uniquely determined and is said to be maximal in Ref.~\cite{K3}, satisfying the conditions shown in Lemma~\ref{lem:koashi_imoto_decomposition_set}.
For obtaining the Koashi-Imoto decomposition of a given bipartite state $\psi^{RA}$,
whether the decomposition in the form of Eq.~\eqref{eq:decomposition_form} is maximal can also be checked by calculating operators on $\mathcal{H}^R\otimes\mathcal{H}^{a_j^L}\otimes\mathcal{H}^{a_j^R}$ for all $j$
\begin{equation}
  \label{eq:product_operator}
  \begin{split}
    &\psi^{R a_j^L a_j^R}\\
    &\coloneqq\left(\mathbbm{1}^R\otimes\Pi^{a_j^L}\otimes\Pi^{a_j^R}\right)\psi^{RA}\left(\mathbbm{1}^R\otimes\Pi^{a_j^L}\otimes\Pi^{a_j^R}\right),
  \end{split}
\end{equation}
and if the decomposition is maximal, each of these operators is a tensor product of operators of $\mathcal{H}^R\otimes\mathcal{H}^{a_j^R}$ and $\mathcal{H}^{a_j^L}$.

In the following, we discuss how to perform the $L$-decomposing procedure and the $R$-combining procedure in our case of the Koashi-Imoto decomposition of $S_\psi^{A|R}\coloneqq\left\{\psi^A\left(\Lambda^R\right):\Lambda^R\geqq 0\right\}$ defined as Eq.~\eqref{eq:psi_lambda} in the main text.

\textit{The $L$-decomposing procedure}: (See also Lemma~3 in Ref.~\cite{K3}.)
Given an intermediate decomposition in the form of Eq.~\eqref{eq:decomposition_form},
the $L$-decomposing procedure aims to decompose a Hilbert space $\mathcal{H}^{a_{j_0}^L}$ in this given decomposition into two subspaces,
so that the decomposition is refined as
\[
  \mathcal{H}^{a_{j_0}^L}\otimes\mathcal{H}^{a_{j_0}^R}=\left(\mathcal{H}_{+}^{a_{j_0}^L}\otimes\mathcal{H}^{a_{j_0}^R}\right)\oplus\left(\mathcal{H}_{-}^{a_{j_0}^L}\otimes\mathcal{H}^{\tilde{a}_{j_0}^R}\right),
\]
where the right-hand side represents subspaces in a refined decomposition satisfying
\[
  \begin{split}
    \mathcal{H}^{a_{j_0}^L}&=\mathcal{H}_{+}^{a_{j_0}^L}\oplus\mathcal{H}_{-}^{a_{j_0}^L}.
  \end{split}
\]
For the Koashi-Imoto decomposition of $S_\psi^{A|R}$,
this refinement is achieved in the following way.
\begin{enumerate}[{Step $L$}-1:]
  \item Find $j_0\in\left\{0,\ldots,J-1\right\}$, $\Ket{a}\in\mathcal{H}^{a_{j_0}^R}$, $\Ket{b}\in\mathcal{H}^{a_{j_0}^R}$, and $\Lambda^R\geqq 0$ such that for any $c\geqq 0$
    \[
      \rho\neq c\rho^\prime,
    \]
    where $\rho$ and $\rho^\prime$ are operators on $\mathcal{H}^{a_{j_0}^L}$ defined as
    \[
      \begin{split}
        \rho&\coloneqq\left(\Pi^{a_{j_0}^L}\otimes\Bra{a}^{a_{j_0}^R}\right)\psi^A\left(\Lambda^R\right)\left(\Pi^{a_{j_0}^L}\otimes\Ket{a}^{a_{j_0}^R}\right),\\
        \rho^\prime&\coloneqq\left(\Pi^{a_{j_0}^L}\otimes\Bra{b}^{a_{j_0}^R}\right)\psi^A\left(\mathbbm{1}^R\right)\left(\Pi^{a_{j_0}^L}\otimes\Ket{b}^{a_{j_0}^R}\right).
      \end{split}
    \]
  \item Calculate the spectral decomposition of an operator on $\mathcal{H}^{a_{j_0}^L}$
    \[
      \frac{\rho}{\tr\rho}-\frac{\rho^\prime}{\tr\rho^\prime}=\sum_l \lambda_l\Ket{l}\Bra{l}.
    \]
    Using the subspaces spanned by eigenvectors of this operator corresponding to the positive eigenvalues and the non-positive eigenvalues, decompose $\mathcal{H}^{a_{j_0}^L}$ into
    \[
      \mathcal{H}^{a_{j_0}^L}=\mathcal{H}_{+}^{a_{j_0}^L}\oplus\mathcal{H}_{-}^{a_{j_0}^L},
    \]
    where the subspaces on the right-hand side are defined as
    \[
      \begin{split}
        \mathcal{H}_{+}^{a_{j_0}^L}&\coloneqq \spn \left\{\Ket{l}\in\mathcal{H}^{a_{j_0}^L}:\lambda_l>0\right\},\\
        \mathcal{H}_{-}^{a_{j_0}^L}&\coloneqq\spn\left\{\Ket{l}\in\mathcal{H}^{a_{j_0}^L}:\lambda_l\leqq 0 \right\}.
      \end{split}
    \]
    Note that $\mathcal{H}_{+}^{a_{j_0}^L}$ and $\mathcal{H}_{-}^{a_{j_0}^L}$ are nonzero subspaces.
  \item Define a refined decomposition as
    \[
      \begin{split}
        \mathcal{H}^A&=\bigoplus_{j=0}^{\tilde{J}-1}\mathcal{H}^{\tilde{a}_j^L}\otimes\mathcal{H}^{\tilde{a}_j^R},\\
        \tilde{J}&\coloneqq J+1,\\
        \mathcal{H}^{\tilde{a}_j^L}&\coloneqq\begin{cases}
          \mathcal{H}^{a_j^L}&\textup{if } 0\leqq j\leqq j_0-1,\\
          \mathcal{H}^{a_{j-1}^L}&\textup{if } j_0\leqq j\leqq J-2,\\
          \mathcal{H}_{+}^{a_{j_0}^L}&\textup{if } j=J-1,\\
          \mathcal{H}_{-}^{a_{j_0}^L}&\textup{if } j=J,\\
        \end{cases}\\
        \mathcal{H}^{\tilde{a}_j^R}&\coloneqq\begin{cases}
          \mathcal{H}^{a_j^R}&\textup{if } 0\leqq j\leqq j_0-1,\\
          \mathcal{H}^{a_{j-1}^R}&\textup{if } j_0\leqq j\leqq J-2,\\
          \mathcal{H}^{a_{j_0}^R}&\textup{if } j=J-1,\, J.
        \end{cases}
      \end{split}
    \]
\end{enumerate}

\textit{The $R$-combining procedure}: (See also Lemma~4 in Ref.~\cite{K3}.)
Given an intermediate decomposition in the form of Eq.~\eqref{eq:decomposition_form},
the $R$-combining procedure aims to combine two different Hilbert spaces $\mathcal{H}^{a_{j_0}^R}$ and $\mathcal{H}^{a_{j_1}^R}$ in this given decomposition into one,
so that the decomposition is refined as
\[
  \begin{split}
    &\left(\mathcal{H}^{a_{j_0}^L}\otimes\mathcal{H}^{a_{j_0}^R}\right)\oplus\left(\mathcal{H}^{a_{j_1}^L}\otimes\mathcal{H}^{a_{j_1}^R}\right)\\
    &=\left(\mathcal{H}^{a_{j_0\cap j_1}^L}\otimes\left(\mathcal{H}^{a_{j_0}^R}\oplus\mathcal{H}^{a_{j_1}^R}\right)\right)\\
    &\quad\oplus\left(\mathcal{H}_\perp^{a_{j_0}^L}\otimes\mathcal{H}^{a_{j_0}^R}\right)\oplus\left(\mathcal{H}_\perp^{a_{j_1}^L}\otimes\mathcal{H}^{a_{j_1}^R}\right),
  \end{split}
\]
where the right-hand side represents subspaces in a refined decomposition satisfying
\[
  \begin{split}
    \mathcal{H}^{a_{j_0}^L}&=\mathcal{H}^{a_{j_0\cap j_1}^L}\oplus\mathcal{H}_\perp^{a_{j_0}^L},\\
    \mathcal{H}^{a_{j_1}^L}&=\mathcal{H}^{a_{j_0\cap j_1}^L}\oplus\mathcal{H}_\perp^{a_{j_1}^L}.\\
  \end{split}
\]
For the Koashi-Imoto decomposition of $S_\psi^{A|R}$,
this refinement is achieved in the following way.
\begin{enumerate}[{Step $R$}-1:]
  \item Find $j_0\in\left\{0,\ldots,J-1\right\}$, $j_1\in\left\{0,\ldots,J-1\right\}$, $\Ket{a}\in\mathcal{H}^{a_{j_0}^R}$, $\Ket{b}\in\mathcal{H}^{a_{j_1}^R}$, and $\Lambda^R\geqq 0$ such that $j_0 < j_1$ and
    \[
      \begin{split}
        &\supp\left(\left(\Pi^{a_{j_0}^L}\otimes\Bra{a}^{a_{j_0}^R}\right)\psi^A\left(\Lambda^R\right)\left(\Pi^{a_{j_0}^L}\otimes\Ket{a}^{a_{j_0}^R}\right)\right)\\
        &\quad =\mathcal{H}^{a_{j_0}^L},\\
        &\supp\left(\left(\Pi^{a_{j_1}^L}\otimes\Bra{b}^{a_{j_1}^R}\right)\psi^A\left(\Lambda^R\right)\left(\Pi^{a_{j_1}^L}\otimes\Ket{b}^{a_{j_1}^R}\right)\right)\\
        &\quad=\mathcal{H}^{a_{j_1}^L},\\
        &\sigma\neq\boldsymbol{0},
      \end{split}
    \]
    where $\supp(\cdots)$ represents the support, $\boldsymbol{0}$ is the zero operator, and $\sigma$ is an operator from $\mathcal{H}^{a_{j_0}^L}$ to $\mathcal{H}^{a_{j_1}^L}$ defined as
    \[
      \sigma\coloneqq\left(\Pi^{a_{j_1}^L}\otimes\Bra{b}^{a_{j_1}^R}\right)\psi^A\left(\Lambda^R\right)\left(\Pi^{a_{j_0}^L}\otimes\Ket{a}^{a_{j_0}^R}\right).
    \]
  \item Calculate the singular value decomposition of $\sigma$
    \[
      \sigma=\sum_{l=0}^{R-1} \sigma_l\Ket{l}^{a_{j_1}^L}\Bra{l}^{a_{j_0}^L},
    \]
    where $R\coloneqq\rank\sigma$, and $\sigma_0,\ldots,\sigma_{R-1}$ are the positive singular values.
    Using the subspace spanned by the singular vectors $\left\{\Ket{0},\ldots,\Ket{R-1}\right\}$ of $\sigma$ corresponding to the positive singular values, decompose $\mathcal{H}^{a_{j_0}^L}$ and $\mathcal{H}^{a_{j_1}^L}$ into
    \[
      \begin{split}
        \mathcal{H}^{a_{j_0}^L}&=\mathcal{H}^{a_{j_0\cap j_1}^L}\oplus{\mathcal{H}_\perp^{a_{j_0}^L}},\\
        \mathcal{H}^{a_{j_1}^L}&=\mathcal{H}^{a_{j_0\cap j_1}^L}\oplus{\mathcal{H}_\perp^{a_{j_1}^L}},
      \end{split}
    \]
    where the subspaces on the right-hand side are defined as
    \[
      \begin{split}
        \mathcal{H}^{a_{j_0\cap j_1}^L}&\coloneqq\spn\left\{\Ket{0},\ldots,\Ket{R-1}\right\},\\
        \mathcal{H}_\perp^{a_{j_0}^L}&\coloneqq\supp\left(\Pi^{a_{j_0}^L}-\sum_{l=0}^{R-1}\Ket{l}\Bra{l}^{a_{j_0}^L}\right),\\
        \mathcal{H}_\perp^{a_{j_1}^L}&\coloneqq\supp\left(\Pi^{a_{j_1}^L}-\sum_{l=0}^{R-1}\Ket{l}\Bra{l}^{a_{j_1}^L}\right).
      \end{split}
    \]
    Note that $\mathcal{H}_\perp^{a_{j_0}^L}$ and $\mathcal{H}_\perp^{a_{j_0}^L}$ may be zero, and define flags indicating whether $\mathcal{H}_\perp^{a_{j_0}^L}$ and $\mathcal{H}_\perp^{a_{j_0}^L}$ are zero as
    \[
      \begin{split}
        s_{j_0}&\coloneqq\begin{cases}
          0&\textup{if } \mathcal{H}_\perp^{a_{j_0}^L}=\left\{\boldsymbol{0}\right\},\\
          1&\textup{otherwise},
        \end{cases}\\
        s_{j_1}&\coloneqq\begin{cases}
          0&\textup{if } \mathcal{H}_\perp^{a_{j_1}^L}=\left\{\boldsymbol{0}\right\},\\
          1&\textup{otherwise}.
        \end{cases}
      \end{split}
    \]
  \item Define a refined decomposition as
    \[
      \begin{split}
        \mathcal{H}^A&=\bigoplus_{j=0}^{\tilde{J}-1}\mathcal{H}^{\tilde{a}_j^L}\otimes\mathcal{H}^{\tilde{a}_j^R},\\
        \tilde{J}&\coloneqq J-1+s_{j_0}+s_{j_1},\\
        \mathcal{H}^{\tilde{a}_j^L}&\coloneqq\begin{cases}
          \mathcal{H}^{a_j^L}&\textup{if } 0\leqq j\leqq j_0-1,\\
          \mathcal{H}^{a_{j+1}^L}&\textup{if } j_0\leqq j\leqq j_1-2,\\
          \mathcal{H}^{a_{j+2}^L}&\textup{if } j_1-1\leqq j\leqq J-3,\\
          \mathcal{H}^{a_{j_0\cap j_1}^L}&\textup{if } j=J-2,\\
          \mathcal{H}_\perp^{a_{j_0}^L}&\textup{if } j = J-2+s_{j_0}\\
                                       &\textup{and }s_{j_0}=1,\\
          \mathcal{H}_\perp^{a_{j_1}^L}&\textup{if } j = J-2+s_{j_0}+s_{j_1}\\
                                       &\textup{and }s_{j_1}=1,\\
        \end{cases}\\
        \mathcal{H}^{\tilde{a}_j^R}&\coloneqq\begin{cases}
          \mathcal{H}^{a_j^R}&\textup{if } 0\leqq j\leqq j_0-1,\\
          \mathcal{H}^{a_{j+1}^R}&\textup{if } j_0\leqq j\leqq j_1-2,\\
          \mathcal{H}^{a_{j+2}^R}&\textup{if } j_1-1\leqq j\leqq J-3,\\
          \mathcal{H}^{a_{j_0}^R}\oplus\mathcal{H}^{a_{j_1}^R}&\textup{if } j=J-2,\\
          \mathcal{H}^{a_{j_0}^R}&\textup{if } j = J-2+s_{j_0}\\
                                 &\textup{and } s_{j_0}=1,\\
          \mathcal{H}^{a_{j_1}^R}&\textup{if } j = J-2+s_{j_0}+s_{j_1}\\
                                 &\textup{and } s_{j_1}=1.
        \end{cases} 
      \end{split}
    \]
\end{enumerate}

In the following, we demonstrate how to obtain the Koashi-Imoto decomposition of a tripartite pure state using the above algorithm.

\begin{example}
  \textit{Koashi-Imoto decomposition of a tripartite pure state.}
Consider a tripartite pure state
\[
  \begin{split}
    &\Ket{\psi}^{RAB}\\
    &\coloneqq\frac{1}{2\sqrt{2}}{\left(\Ket{0}^{R}\otimes\Ket{0}^{A_1}+\Ket{1}^{R}\otimes\Ket{1}^{A_1}\right)}\\
    &\qquad\otimes{\left(\Ket{0}^{A_2}\otimes\Ket{0}^{B}+\Ket{1}^{A_2}\otimes\Ket{1}^{B}\right)}\\
    &\quad +\frac{1}{\sqrt{2}}\Ket{2}^{R}\otimes\Ket{2}^{A_1}\otimes\Ket{0}^{A_2}\otimes\Ket{2}^{B},
  \end{split}
\]
where $\mathcal{H}^R$ is of $3$ dimension, $\mathcal{H}^A=\mathcal{H}^{A_1}\otimes\mathcal{H}^{A_2}$ of $3\times 2 = 6$ dimension, and $\mathcal{H}^{B}$ of $3$ dimension.
The Koashi-Imoto decomposition can be algorithmically obtained as follows,
where the order of subspaces in intermediate decompositions is sorted for readability.
\begin{enumerate}[{Step} 1:]
  \item Initially, regard $\mathcal{H}^A$ as
    \begin{equation}
      \label{eq:1}
      \mathcal{H}^A=\mathcal{H}^{a_0^L}.
    \end{equation}
  \item Apply the $L$-decomposing procedure to the intermediate decomposition given by Eq.~\eqref{eq:1}, where $j_0=0$, $\Ket{a}=1$, $\Ket{b}=1$, and $\Lambda^R=\Ket{0}\Bra{0}$, and $\mathcal{H}^A$ is decomposed into
    \begin{equation}
      \label{eq:2}
      \mathcal{H}^A=\mathcal{H}^{a_0^L}\oplus\mathcal{H}^{a_1^L},
    \end{equation}
    where $\dim\mathcal{H}^{a_0^R}=\dim\mathcal{H}^{a_1^R}=1$ and
    \[
      \begin{split}
        \mathcal{H}^{a_0^L}=\spn\Big\{&\Ket{0}^{A_1}\otimes\Ket{0}^{A_2},\Ket{0}^{A_1}\otimes\Ket{1}^{A_2}\Big\},\\
        \mathcal{H}^{a_1^L}=\spn\Big\{&\Ket{1}^{A_1}\otimes\Ket{0}^{A_2},\Ket{1}^{A_1}\otimes\Ket{1}^{A_2},\\
                                      &\Ket{2}^{A_1}\otimes\Ket{0}^{A_2},\Ket{2}^{A_1}\otimes\Ket{1}^{A_2}\Big\},\\
      \end{split}
    \]
  \item Apply the $L$-decomposing procedure to the intermediate decomposition given by Eq.~\eqref{eq:2}, where $j_0=1$, $\Ket{a}=1$, $\Ket{b}=1$, and $\Lambda^R=\Ket{1}\Bra{1}$, and $\mathcal{H}^A$ is decomposed into
    \begin{equation}
      \label{eq:3}
      \mathcal{H}^A=\mathcal{H}^{a_0^L}\oplus\mathcal{H}^{a_1^L}\oplus\mathcal{H}^{a_2^L},
    \end{equation}
    where $\dim\mathcal{H}^{a_0^R}=\dim\mathcal{H}^{a_1^R}=\dim\mathcal{H}^{a_2^R}=1$ and
    \[
      \begin{split}
        \mathcal{H}^{a_0^L}=\spn\Big\{&\Ket{0}^{A_1}\otimes\Ket{0}^{A_2},\Ket{0}^{A_1}\otimes\Ket{1}^{A_2}\Big\},\\
        \mathcal{H}^{a_1^L}=\spn\Big\{&\Ket{1}^{A_1}\otimes\Ket{0}^{A_2},\Ket{1}^{A_1}\otimes\Ket{1}^{A_2}\Big\},\\
        \mathcal{H}^{a_2^L}=\spn\Big\{&\Ket{2}^{A_1}\otimes\Ket{0}^{A_2},\Ket{2}^{A_1}\otimes\Ket{1}^{A_2}\Big\}.
      \end{split}
    \]
  \item Apply the $R$-combining procedure to the intermediate decomposition given by Eq.~\eqref{eq:3}, where $j_0=0$, $j_1=1$, $\Ket{a}=1$, $\Ket{b}=1$, and $\Lambda^R=\Ket{0}\Bra{0}+\Ket{0}\Bra{1}+\Ket{1}\Bra{0}+\Ket{1}\Bra{1}$, and $\mathcal{H}^A$ is decomposed into
    \begin{equation}
      \label{eq:4}
      \mathcal{H}^A=\left(\mathcal{H}^{a_0^L}\otimes\mathcal{H}^{a_0^R}\right)\oplus\mathcal{H}^{a_1^L},
    \end{equation}
    where $\dim\mathcal{H}^{a_1^R}=1$ and
    \[
      \begin{split}
        \mathcal{H}^{a_0^L}=\spn\Big\{&\Ket{0}^{A_2},\Ket{1}^{A_2}\Big\},\\
        \mathcal{H}^{a_0^R}=\spn\Big\{&\Ket{0}^{A_1},\Ket{1}^{A_1}\Big\},\\
        \mathcal{H}^{a_1^L}=\spn\Big\{&\Ket{2}^{A_1}\otimes\Ket{0}^{A_2},\Ket{2}^{A_1}\otimes\Ket{1}^{A_2}\Big\}.
      \end{split}
    \]
  \item Terminate the algorithm, since for each $j$, the operator $\psi^{R a_j^L a_j^R}$ defined as Eq.~\eqref{eq:product_operator} is a tensor product of operators of $\mathcal{H}^R\otimes\mathcal{H}^{a_j^R}$ and $\mathcal{H}^{a_j^L}$, and hence, the decomposition in Eq.~\eqref{eq:4} is maximal. In this case, $\Ket{\psi}^{RAB}$ is decomposed into
    \[
      \begin{split}
        &\Ket{\psi}^{RAB}\\
        &=\frac{1}{2\sqrt{2}}{\left(\Ket{0}^{R}\otimes\Ket{0}^{a_0^R}+\Ket{1}^{R}\otimes\Ket{1}^{a_0^R}\right)}\\
        &\qquad\otimes{\left(\Ket{0}^{a_0^L}\otimes\Ket{0}^{b_0^L}+\Ket{1}^{a_0^L}\otimes\Ket{1}^{b_0^L}\right)}\\
        &\quad \oplus\frac{1}{\sqrt{2}}\left(\Ket{2}^{R}\otimes{\left(\Ket{2}\otimes\Ket{0}\right)}^{a_1^L}\otimes\Ket{2}^{b_1^L}\right).
      \end{split}
    \]
\end{enumerate}
\end{example}

\bibliographystyle{IEEEtran}
\bibliography{citation_bibtex}

\end{document}